\documentclass{article}
\usepackage{babel}    
\usepackage[margin=2.5cm]{geometry}
\usepackage{graphicx}
\usepackage{lipsum}
\usepackage[dvipsnames]{xcolor}
\usepackage{epsfig}
\usepackage{float}
\usepackage{cite}
\usepackage{epstopdf}
\usepackage{caption}
\usepackage{subcaption}
\usepackage{amsmath}
\usepackage{graphicx}
\usepackage{amsmath}

\title{Gravitational collapse in pure Gauss- Bonnet theory}
\author{Akshay Kumar\footnote{akshay.relativity@gmail.com} \,\,and Ayan Chatterjee\footnote{ayan.theory@gmail.com}\,\\
Department of Physics and Astronomical Science,\\
Central University of Himachal Pradesh,\\
Dharamshala- 176215, India. \,
\vspace{0.2cm}\\
Suresh C. Jaryal\footnote{suresh.fifthd@gmail.com} \\
Department of Physics and Photonics Science, \\
National Institute of Technology Hamirpur,\\
Hamirpur, Himachal Pradesh- 177005, India. }
\begin{document}
\date{}
\maketitle
\begin{abstract}
    In this paper, we study the gravitational collapse of matter fields, which include dust, perfect fluids as well as fluids admitting bulk and shear viscosity. The initial conditions on these matter fields have been kept to be quite general: the initial velocity profile of the matter is taken to include both the bound and the marginally bound models, while the density profile of the initial matter configuration is assumed to have physically admissible portrayal, and smooth falloffs. We determine, under these general conditions, the time of formation of the central singularity and the formation and evolution of black hole horizons, depicted here in terms of quasilocal marginally trapped surfaces. Our study shows that under these general conditions, the central singularity remains hidden from the asymptotic observer. 
\end{abstract}
\vspace{0.5 cm}
PACS:
\section{Introduction}
It is a matter of interest to understand the final outcome of a continuous gravitational collapse  of matter fields in Einstein's general relativity (GR) and in the alternative models including higher order gravity theories such as the Lovelock theory. The singularity theorems propose that, under reasonable conditions, the occurrence of a spacetime singularity in GR is an inevitable consequence of such a gravitational collapse of any massive star \cite{Hawking_Ellis, Wald}. The cosmic censorship conjecture ensures that these singularities always remain hidden within the event horizon (EH), at least in the context of GR \cite{Penrose:1964wq, Penrose:1969pc}. Indeed, the pioneering work of Oppenheimer, Snyder\cite{OS}, and Datt\cite{SD} demonstrates that the collapse of a spherical homogeneous dust ball, not only forms a singularity but also that it remains causally disconnected from asymptotic observers. Although, studies carried out so far has contributed to a better understanding the censorship conjecture, yet only some of the simplest collapsing scenarios like the dust and certain fluid matter with reasonable equations of state, and anisotropic collapse models have been explored due to mathematical complexity involved under more relaxed assumptions \cite{joshi,Joshi:2012mk, Clarke, Landau_Lifshitz, Ori}. Most studies show that depending on the initial conditions, endless gravitational collapse of massive star can have two possibilities: first, the central singularity is hidden behind an event horizon which results the formation of black hole (BH), or the second,  the singularity is potentially visible or is naked (NS)\cite{Ori, joshi,Joshi:2012mk, Clarke, Shapiro:1991zza, Dwivedi:1992fh, Ghosh:2001fb, Harada:2001nj, Joshi:2004tb}. The visibility of the central singularity raises the question on the validity of the cosmic censorship conjecture. 
Even though Einstein's general relativity is an exceptionally reliable theory of gravity yet it has a defined regime of validity. It is believed that modifications to the GR can resolve the issues affecting GR \cite{Hawking:1972qk, Sotiriou:2011dz, Brown:2018hym}. The motivation behind this growing interest is to understand the open unresolved questions in the context of these modifications to GR or higher order gravity theories. For example, one may ask the following questions: 
(i) what are the effects of higher order correction terms (modification to GR), on the formation of  singularities, and whether these singularities (central as well non central) are hidden behind event horizon or if they are potentially visible to the faraway observer, (ii) is it possible to demonstrate validity of cosmic censorship conjecture, and  (iii) if these modifications can offer non singular solutions that avoid infinite curvature.\\

The modification to GR can be achieved by incorporating the higher curvature terms in the Einstein-Hilbert action. The Lovelock- Lanczos theories \cite{Lovelock:1971yv, Lanczos:1938sf} are the natural choice for studying corrections to GR at higher dimensions. They are the natural choice in the sense that they provide a homogeneous polynomial of degree $N$ in Riemann curvature these theories are ghost-free and yield second-order gravitational equations. More specifically, for $N=0,1,2, \cdots$, it gives cosmological constant, GR and Gauss-Bonnet (GB) gravity, respectively. The pure Lovelock action is obtained when only a particular $N$ is considered. For example, GR is a pure Lovelock term for $N=1$. It has been argued from the universality of kinematic property of Lovelock theories that non-trivial vacuum solution does not exist in all critical odd $n=2N+1$ dimensions \cite{Dadhich:2012cv, Dadhich:2015lra}, and this is a potential reason for considering the pure Lovelock theories as better candidate for higher dimensional gravity theories. The pure Lovelock theories focus on a single $N^{th}$ order term and unlike general Lovelock theories, eliminate the coupling constant value dependence, and therefore have several appealing and important properties \cite{Dadhich:2012cv, Dadhich:2015lra, Camanho:2015hea, Kothawala:2009kc, Chakraborty:2014rga, Chakraborty:2016qbw, Gannouji:2019gnb, Dadhich:2016fku, Dadhich:2016wtb,  Shaymatov:2020byu, Dadhich:2017zdi, Molina:2016xeu, Singha:2023lum, Dadhich:2013bya}.\\

In this paper, we deal with the gravitational collapse of inhomogeneous matter in pure GB theory for the following types of matter fields: dust, perfect and viscous matter fields (anisotropic collapse). By assuming these as the initial matter profiles, we determine the radius of the collapsing shells as a function of proper time and shell coordinate, and hence, we can determine the exact solutions of the pure GB field equations. These solutions can be used to determine the formation of the central singularity, the spherically symmetric trapped surfaces and the evolution of boundary of trapped surface with time as more matter fall in. In general, it has been observed that the formation of singularity, trapped surfaces and their evolution mainly depends on theory and the initial density and their velocity profiles of the collapsing matter\cite{Booth:2005ng,Chatterjee:2020khj,Chatterjee:2021zre,Jaryal:2022rzd,Chatterjee:2024egb}.
Since we are interested to study the effects of GB term, we shall use a wide range of physically realistic initial matter- density distribution for the collapsing configuration in the pure GB gravity. A similar study  for GR and EGB is in \cite{Chatterjee:2020khj,Chatterjee:2021zre, Jaryal:2022rzd,Chatterjee:2024egb}.
In each of these density profiles, we track time development of spacetime singularity and the motion of the collapsing shells. We shall be able to precisely locate the formation of spherically trapped surface also known as Marginally trapped surfaces (MTT) and evolution of their boundary (black hole horizon) as more matter shells starts it falls in. Note that  marginally trapped tubes (MTTs) are fundamental for studying the dynamics of the horizon and the visibility of singularities in various gravitational theories \cite{ 
Ashtekar:2004cn, Booth:2005ng, Chatterjee:2020khj, Chatterjee:2021zre, Jaryal:2022rzd, Chatterjee:2024egb}. 
In general, these horizons are defined through event horizon \cite{Hawking_Ellis, Wald}, however due to teleological nature of event horizon, this definition suffers from conceptual contradiction, hence such quasilocal definitions of horizon are useful \cite{Ashtekar:2004cn}. Here, the horizon is $(n-1)$ dimensional hypersurface foliated by $(n-2)$ spacelike spheres along which outgoing null normal expansion is $\theta_{(l)}=0$ and ingoing null normal expansion is $\theta_{(n)}<0$. These MTTs are found useful for studying different properties of black holes both classically and quantum \cite{ 
Ashtekar:2004cn, Booth:2005ng, Chatterjee:2020khj, Chatterjee:2021zre, Jaryal:2022rzd, Chatterjee:2024egb, Raviteja:2021dgy, gutti2}. 
During the collapse, MTTs can have signature null, and spacelike. This signature governs the properties of the horizon as, for null signature, the black hole is at equilibrium and named an isolated horizon (IH), for spacelike signature the black hole is admitting matter and growing and so-called dynamical horizon (DH) \cite{Ashtekar:2004cn, Booth:2005ng}. While significant progress has been made in GR and EGB gravity \cite{Chatterjee:2020khj, Chatterjee:2021zre, Chatterjee:2024egb, Jaryal:2022rzd}, the pure GB gravity is less explored in the context of MTTs\cite{Dadhich:2013bya, Maeda:2005ci, Cai:2006pq, Garraffo:2008hu, Mukherjee:2020lld, Mukherjee:2021erg, Dialektopoulos:2023qda}. Understanding the role of MTTs in pure GB gravity can shed light on the dimensionality dependence, initial conditions, and clear up correspondence of results with GR. This study aims to address this gap by extending the formalism of MTTs to pure GB gravity and comparing its predictions with GR results. \\

The paper is structured as follows: In the next section, we have discussed the field equations in pure GB gravity for spherical symmetric spacetime metric in $n$ dimensions. In section $3$ we have discussed the pressureless collapse scenario, and specifically the inhomogeneous collapse of marginally bound and bound fluids. In section $4$, the viscous matter fields and perfect fluids are discussed. The work is concluded in the section $5$.
\section{The formalism of spherical gravitational collapse}
The action for pure Gauss-Bonnet (GB) gravity is given by,
\begin{equation}\label{GBaction }
    S=\int{d^{n}x\sqrt{-g}L_{GB}+S_{matter}},
\end{equation}
where $g$ is the determinant of the metric $g_{\mu\nu}$ and the GB Lagrangian($L_{GB}$) is given by,
\begin{equation} L_{GB}=R^{2}-4R_{\mu\nu}R^{\mu\nu}+R_{\mu\nu\sigma\delta}R^{\mu\nu\sigma\delta},
\end{equation}
where $R$ is Ricci scalar, $R_{\mu\nu}$ denotes the Ricci tensor and $R_{\mu\nu\sigma\delta}$ is the Riemann tensor. The variation of action in eqn. (\ref{GBaction }) leads to the following field equations,
\begin{equation}
    H_{\mu\nu}=T_{\mu\nu},
\end{equation}
where $H_{\mu\nu}$ is the GB tensor and $T_{\mu\nu}$ is the energy momentum tensor, the GB tensor is define as,
\begin{equation}
    H_{\mu\nu}= 2\left[RR_{\mu\nu}-2R_{\mu\lambda}{R^{\lambda}}_{\nu}-2R^{\lambda\sigma}R_{\mu\lambda\nu\sigma}+{R_{\mu}}^{\lambda\sigma\delta}R_{\nu\lambda\sigma\delta}\right]-\frac{1}{2}g_{\mu\nu}L_{GB}.
\end{equation}
We define the metric for $n$-dimensional spherical symmetric spacetime as,
 \begin{equation}\label{metric}
  ds^2= e^{2\alpha(t,r)}dt^2+e^{2\beta(t,r)}dr^2+R^2(t,r)d\Omega^2_{n-2}\,\,,
  \end{equation}
where,$$ d\Omega^2_{n-2}= \sum_{i=1}^{n-2} \left[ \prod_{j}^{i-1} \sin^{2} (\theta^{j}) \right](d\theta^{i})^{2}.$$
The field equations will be solved for the energy momentum tensor $T_{\mu\nu}$ of a spherical ball with the following form,
  \begin{equation}
          T_{\mu\nu}=(p_{t}+\rho)\,u_{\mu}\,u_{\nu}+p_{t}\,g_{\mu\nu}+(p_{r}-p_{t})\,X_{\mu}\,X_{\nu}-2\,\eta\,\sigma_{ab}-\zeta\,\theta\,h_{\mu\nu}\,\,,
      \end{equation}
      where $\eta$ and $\zeta$ are the coefficients of shear and bulk viscosity, $X^{\mu}$ is a unit space-like vector tangential to the space like section orthogonal to to $u^{\mu}$ respectively satisfying $X_{\mu}X^{\mu}=1$. The quantities $\sigma_{\mu\nu}$, $\theta$, and $h_{\mu\nu}$ are shear, expansion and projection tensor and, $\rho$, $p_{t}$ and $p_{r}$ are the energy density and tangential and radial components of pressure respectively. The expression for these quantities are
      \begin{eqnarray}
      \theta&=&\nabla_{\mu}u^{\mu},~~~~
         {h^{\mu\nu}}=({g^{\mu\nu}}+u^{\mu}\,u^{\nu}), ~~~~
        X^{\mu}=e^{-2\beta(t,r)}\delta^{\mu}_{r}\\
      \sigma^{\mu\nu}&=&(1/2)(h^{\mu\gamma}\,\nabla_{\gamma}u^{\nu}+h^{\nu\gamma}\,\nabla_{\gamma}u^{\mu})-[1/(n-1)]\,\theta\, h^{\mu\nu}
     \end{eqnarray}
The values for these scalars for the above metric is easily determined to be:
      \begin{eqnarray}
          \theta=e^{-\alpha}\,\left(\dot{\beta}+(n-2)\dot{R}/R\right)\,\,  ,
        ~~~~   \sigma=e^{-\alpha}(\dot{\beta}-\dot{R}/R).
      \end{eqnarray}
The non zero components of the energy momentum tensor are given by the following quantities,
          \begin{eqnarray}
              {T^{0}}_{0}=-\rho\,,\hspace{0.5cm}
              {T^{1}}_{1}=p_{r}-\frac{2(n-2)\eta\sigma}{n-1}-\zeta\theta\,,\hspace{0.5cm}  {T^{a}}_{b}=\left[p_{t}+\frac{2\eta\sigma}{n-1}-\zeta\theta\right]{\delta^{a}}_{b}\,,
          \end{eqnarray}
          where $a=b\geq2$. The equation of motion for the metric (\ref{metric}) in the pure GB theory will get the following form,
          \begin{eqnarray}
              \rho(t,r)&=&\frac{(n-2)(n-3)(n-4)}{2}\frac{F'(t,r)}{R^{n-2}R'}\,\, ,\label{H00}\\
              p_{r}(t,r)&=&\frac{2(n-2)\eta\sigma}{(n-1)}+\zeta\theta-\frac{(n-2)(n-3)(n-4)}{2}\frac{\dot{F}(t,r)}{R^{n-2}\dot{R}}\,\, ,\label{H11}\\
              \alpha'&=& -\frac{(n-2)R'}{R}\frac{\left[p_{r}-p_{t}+2\eta\sigma \right]}{\left[\rho+p_{r}-2(n-2)\eta\sigma/(n-1)-\zeta\theta \right]}-\frac{\left[p_{r}-2(n-2)\eta\sigma/(n-1)-\zeta\theta \right]'}{\left[\rho+p_{r}-2(n-2)\eta\sigma/(n-1)-\zeta\theta \right]}  \,\,,\label{bianchi}\\
              2{\dot{R}}'&=&R'\frac{\dot{G}}{G}+\dot{R}\frac{H'}{H}\,\,,\label{H01}\\
              F(t,r)&=&R^{n-5}(1-G+H)^{2}\,\, \label{mass function},
         \end{eqnarray}
         the first two equations (\ref{H00}) and (\ref{H11}) are the $H_{00}$ and $H_{11}$ equations. The third equation (\ref{bianchi}) came from Bianchi identity. The fourth equation (\ref{H01}) is $H_{01}$ equation and the fifth one (\ref{mass function}) is a Misner-Sharp mass function. The superscripts $(.)$ and $(')$ are partial derivatives with respect to $t$ and $r$ respectively. In the equation of mass function (\ref{mass function}) the functions $G(t,r)$ and $H(t,r)$ are defined as, $G(t,r)=e^{-2\beta}{R'}^{2}$ and $H(t,r)=e^{-2\alpha}{\dot{R}}^{2}$. Note that black holes exist for $n\ge 6$.\\

\section{Gravitational collapse pressureless matter}
We shall use above equations to understand the pressureless collapse scenario, where the pressure components $p_{t}=p_{r}=0$ are equal and vanishing and also the viscosity coefficients $\eta$ and $\zeta$ are vanishing. This leads to the following modified equations from (\ref{H00})-(\ref{H01}),
        \begin{eqnarray}
            F^{\prime}(r,t)&=&\frac{2\rho R' R^{n-2}}{(n-2)(n-3)(n-4)}\,\,,\hspace{1.75cm} \dot{F}(r,t)=0\,\,,\label{dust1}\\
            \alpha^{\prime}(t,r)&=&0\,, \hspace{5.0cm} \frac{\dot{R}'}{R'}=\dot{\beta}\,.\label{dust2}
        \end{eqnarray}
        The mass function $\dot{F}=0$, implies that $F=F(r)$, whereas $\alpha'=0$ implies that $\alpha=\alpha(t)$, as the result we can redefine the time coordinate in the metric (\ref{metric}). From the second equation in (\ref{dust2}), we get $R'=e^{\beta(t,r)+h(r)}$ and if we redefine $e^{2h(r)}=1-k(r)$, the metric (\ref{metric}) becomes,
        \begin{equation}
            ds^2=-dt^2+\frac{{R'}^{2}}{1-k(r)}dr^2+R^2(t,r)d\Omega^2_{n-2}\,\,.
        \end{equation}
       This leads to the equation of motion or equation of collapsing shells derived from the mass function equation (\ref{mass function}):
       \begin{equation}\label{eom}
           \dot{R}(r,t)=-\left[\sqrt{\frac{F}{R^{n-5}}}-k(r)\right]^{1/2}
       \end{equation}
 Now the outgoing and ingoing null normals to the $(n-2)$-sphere is given by:
         \begin{eqnarray}
             l^{\mu}&=&\delta_{t}^{\mu}+e^{-\beta(t,r)}\delta_{r}^{\mu}\,\, , ~~~~ n^{\mu}=\frac{1}{2}[\delta_{t}^{\mu}-e^{-\beta(t,r)}\delta_{r}^{\mu}]\,\,\, .
         \end{eqnarray}
          The expressions for $\theta_{l}=h^{\mu\nu}\nabla_{\mu}l_{\nu}$ and $\theta_{n}=h^{\mu\nu}\nabla_{\mu}n_{\nu}$ are as follows:
          \begin{eqnarray}
              \theta_{l}&=&\frac{(n-2)}{R}\left[\dot{R}+\sqrt{1-k(r)}\right]\,\, ,~~~~
              \theta_{n}=\frac{(n-2)}{R}\left[\dot{R}-\sqrt{1-k(r)}\right]\,\, ,
         \end{eqnarray}
         for MTT, $\theta_{l}=0$, so using this condition and equation (\ref{eom}),
         \begin{equation}\label{mtt}
             R_{MTT}=F^{1/(n-5)}\,\, ,  
         \end{equation}
         which is the radius at which MTT will form ($n\geq6$). The parameter $k(r)$ determines the nature of collapse,  $ k(r)=0$ is called marginally bound model, whereas for $k(r)>0$ holds for bound and models.
\subsection{Marginally bound model}  
The equation of motion (\ref{eom}) becomes,
    \begin{equation}\label{eomltb1}
        \dot{R}=-\left[\frac{F(r)}{R^{n-5}}\right]^{1/4}\,\, .
    \end{equation}
    The solution of above equation is the equation of matter shell corresponding to value of $R(t,r)$,
    \begin{equation}\label{tcmb}
        t=-\frac{4}{(n-1)}\frac{R^{(n-1)/4}}{(n-1)}+\frac{4}{(n-1)}\frac{r^{(n-1)/4}}{F^{1/4}}\,\,.
    \end{equation}
    The collapse of shell starts at $t_{i}=0$ and $R(t_{i},r)=r$, we can also find the time for shell to reaching the singularity by putting $R=0$,
    \begin{equation}
        t_{s}=\frac{4}{(n-1)}\frac{r^{(n-1)/4}}{F^{1/4}}\,\,.
    \end{equation}
    The time for the formation of MTT is obtained using equation (\ref{mtt}),
    \begin{equation}\label{tmttmb}
        t_{MTT}=-\frac{4}{(n-1)}\frac{{R_{MTT}}^{(n-1)/4}}{(n-1)}+\frac{4}{(n-1)}\frac{r^{(n-1)/4}}{F^{1/4}}\,\,.
    \end{equation}
\subsection*{Examples}
In the following, the initial densities of the collapsing configuration are taken to be of the following forms:
\begin{eqnarray}
    \label{margdensity1}\rho_{1}(r)&=&\mathcal{N}m_{0}\left(1/{r_{0}}^2\right)^{(n-1)/2}e^{-r^2/{{r_{0}}^2}}\,\,,\\
    \label{margdensity4}\rho_{2}(r)&=&\frac{\alpha}{{r_{0}}^{n-2}}  \left[\pi -\frac{r}{5 r_{0}} \left\{2 \cos ^2\left(5 r/r_{0}\right)+3\right\}\right]\,\,,\\
    \label{margdensity5}\rho_{3}(r)&=&\frac{\alpha  \mu  }{2 \pi ^2 r^{n-2} {r_{0}}^{n-3}}\sin ^2\left(\alpha  r/r_{0}\right)\,\,,
\end{eqnarray}
where $\mathcal{N}=\frac{(n-2) (n-3) (n-4)  }{4 \pi  \Gamma \left((n-1)/2\right)}$ is the normalisation factor.
\begin{figure}[htb]
\begin{subfigure}{.33\textwidth}
\centering
\includegraphics[width=\linewidth]{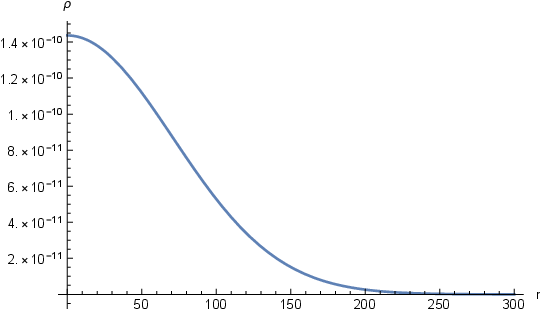}
\caption{}
\end{subfigure}
\begin{subfigure}{.33\textwidth}
\centering
\includegraphics[width=\linewidth]{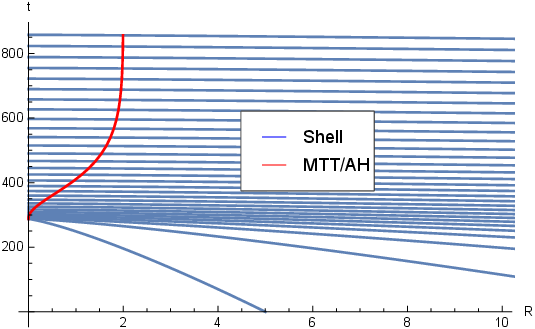}
\caption{}
\end{subfigure}
\begin{subfigure}{.31\textwidth}
\centering
\includegraphics[width=\linewidth]{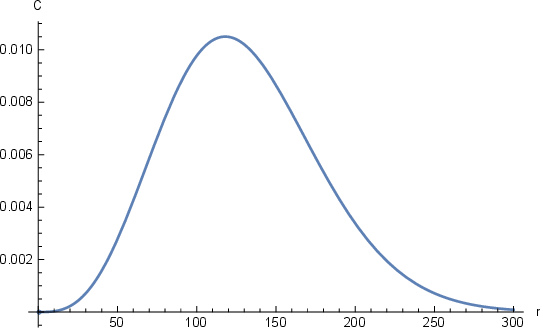}
\caption{}
\end{subfigure}
\caption{The figure (a) shows the density distribution $\rho_{1}$, Eq. \eqref{margdensity1} vs $r$. The Figures (b) shows the time evolution of the collapsing shells Eq. \eqref{tcmb} and MTT Eq \eqref{tmttmb}, and (c) shows the dynamics of $C$ (or nature of MTT) vs $r$ for the marginally bound collapse in $6$- dimensions. It can be seen from Figure (c) that the MTT remains positive throughout the collapse, which implies that its nature is spacelike and it becomes null as the matter stops falling in. 
Here, we have chosen $m_{0}=1.0$ and $r_{0}=100\,m_{0}$. }
\label{figure_marginally1}
\end{figure}
\begin{figure}[htb]
\begin{subfigure}{.33\textwidth}
\centering
\includegraphics[width=\linewidth]{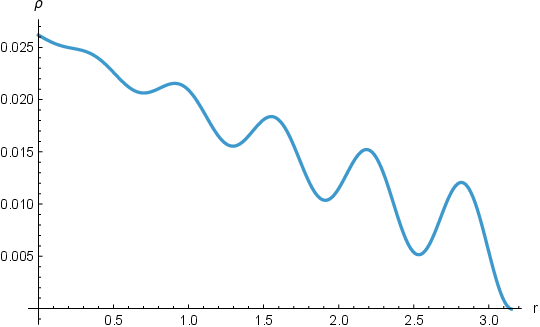}
\caption{}
\end{subfigure}
\begin{subfigure}{.33\textwidth}
\centering
\includegraphics[width=\linewidth]{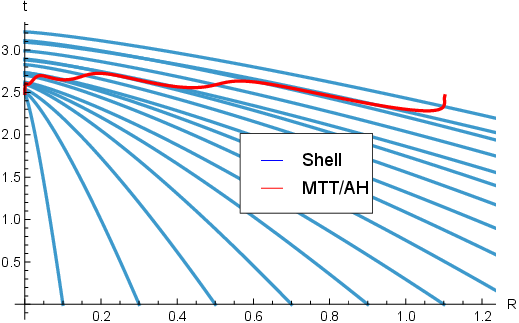}
\caption{}
\end{subfigure}
\begin{subfigure}{.31\textwidth}
\centering
\includegraphics[width=\linewidth]{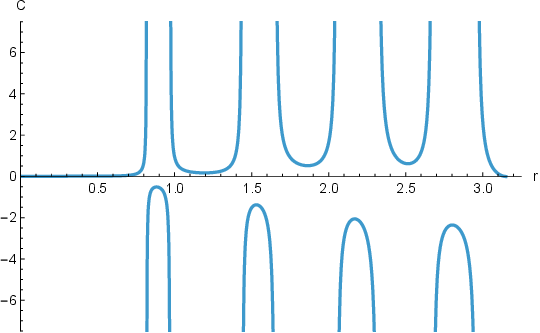}
\caption{}
\end{subfigure}
\caption{The figure (a) shows the irregular density distribution $\rho_{2}$, Eq. \eqref{margdensity4} vs $r$, vanishes at $r=\pi\, r_{0}$. The Figures (b) shows the time evolution of the collapsing shells Eq. \eqref{tcmb} and MTT Eq \eqref{tmttmb}, 
and (c) shows the dynamics of $C$ (or nature of MTT) vs $r$ for the marginally bound collapse in $6$- dimensions. From Figures (b) and (c), it can be seen that as the shells starts to fall in, MTT begins 
 to evolve as a dynamical horizon and then it oscillates between being timelike, non- isolated null and spacelike regions. 
It must be noted that, as the last collapsing shell parameterized by $r=\pi\, r_{0}$ falls in, MTT becomes isolated horizon. Thus, the MTT's evolution as oscillating between different region remains hidden from the farway observer as these regions remains behind the outermost isolated horizon.    
Here, we have chosen $m_{0}=1.0$, $r_{0}=1.0\,m_{0}$ and $\alpha=1/120$.
}
\label{figure_marginally4}
\end{figure}
\begin{figure}[htb]
\begin{subfigure}{.33\textwidth}
\centering
\includegraphics[width=\linewidth]{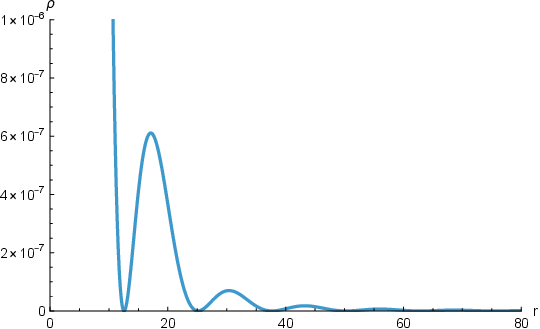}
\caption{}
\end{subfigure}
\begin{subfigure}{.33\textwidth}
\centering
\includegraphics[width=\linewidth]{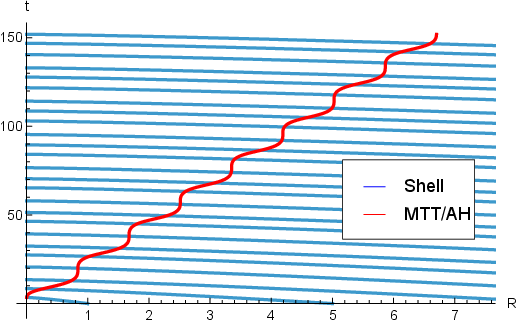}
\caption{}
\end{subfigure}
\begin{subfigure}{.31\textwidth}
\centering
\includegraphics[width=\linewidth]{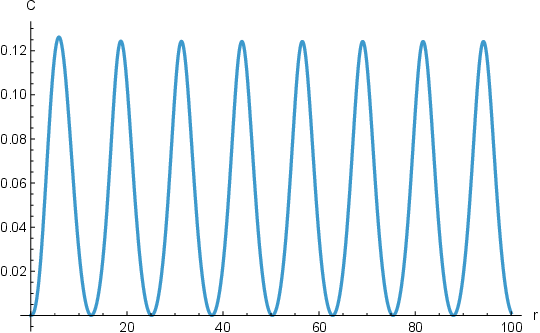}
\caption{}
\end{subfigure}
\caption{The figure (a) shows the density distribution $\rho_{3}$, Eq. \eqref{margdensity5} vs $r$. The Figures (b) shows the time evolution of the collapsing shells Eq. \eqref{tcmb} and MTT Eq. \eqref{tmttmb}, and (c) shows the dynamics of $C$ (or nature of MTT) vs $r$ for the marginally bound collapse in $6$- dimensions. It can be seen from Figure (c) that the MTT remains positive throughout the collapse, which implies that its nature is spacelike and it becomes null as the matter stops falling in. Here, we have chosen $m_{0}=1.0$, $r_{0}=1.0\,m_{0}$, $\alpha=1/4$ and $\mu = (8\pi/5)\,r_{0}$.
}
\label{figure_marginally5}
\end{figure}
\subsection{Bound models}
 For $k(r)>0$ models, the equation (\ref{eom}) can be solved by using parametric choice
    \begin{equation}\label{parametric}
        R=\left[\frac{F}{k^{2}}\cos^{4}\left(\eta/2\right)\right]^{1/(n-5)}\,\,,
    \end{equation}
    Using above equation, the equation of motion (\ref{eom}) becomes,
    \begin{equation}
        dt=\frac{2}{\sqrt{k}}\left[\frac{F}{k^{2}}\cos^{4}\left(\eta/2\right)\right]^{1/(n-5)}d\eta\,\,,
    \end{equation}
    Integration of above equation and re-substituting $\eta$ terms into $R$ gives,
    \begin{eqnarray}\label{tbc}
         t&=&A\left(r^{(n-1)/4}\, _2F_1\left[1/2,(n-3)/2(n-5);(3n-13)/2(n-5);(k^2 r^{n-5}/F)^{1/2}\right]\right.\nonumber\\
         &&\left.-R^{(n-1)/4}\, _2F_1\left[1/2,(n-3)/2(n-5);(3n-13)/2(n-5);(k^2 R^{n-5}/F)^{1/2}\right] \right)\,\,,
    \end{eqnarray}
    where $A=4((n-5)/(n-1))F^{-1/4} k^{1/(n-5)}$. Hereinafter symbol $_2F_1$ represents the $hypergeometric$ function. This equation gives the decrease of radius $R(r,t)$ of the $r$th collapsing shell with time. 
    Again the evolution equation for MTT can be calculated using condition in equation (\ref{mtt}). These expressions are plotted graphically for matter profiles in the following examples:
    \begin{eqnarray}\label{t2mbc}
        t_{MTT}&=&A\left(r^{(n-1)/4}\, _2F_1\left[1/2,(n-3)/2(n-5);(3n-13)/2(n-5);(k^2 r^{n-5}/F)^{1/2}\right]\right.\nonumber\\
         &&\left.-{R_{MTT}}^{(n-1)/4}\, _2F_1\left[1/2,(n-3)/2(n-5);(3n-13)/2(n-5);(k^2 {R_{MTT}}^{n-5}/F)^{1/2}\right] \right)\,.
    \end{eqnarray} 
\subsection*{Examples}
To determine the nature of the collapsing matter, the singularity formation time, and formation of horizon, we use the following examples:
\begin{eqnarray}
    \label{boundeddensity2}\rho_{1}(r)&=&\mathcal{N}_{1}\,m_{0} (10-r) \Theta (10-r)\,\,,\\
    \label{boundeddensity3}\rho_{2}(r)&=&\mathcal{N}_{2}\,m_{0} \left(10-r^2\right) \Theta \left(10-r^2\right)\,\,,
\end{eqnarray}
where $\mathcal{N}_{1}=-\frac{\left(n-n^2\right) \left(n^3-9 n^2+26 n-24\right)}{\pi  2^{n+4} 5^n}$ and $\mathcal{N}_{2}=\frac{n^5-9 n^4+25 n^3-15 n^2-26 n+24}{\pi  2^{\frac{n+11}{2}} 5^{\frac{n+1}{2}}}$ are the normalisation factors. Hereinafter, we have considered, $k=\sqrt{F/r^{n-5}}$.
\begin{figure}
\begin{subfigure}{.5\textwidth}
\centering
\includegraphics[width=\linewidth]{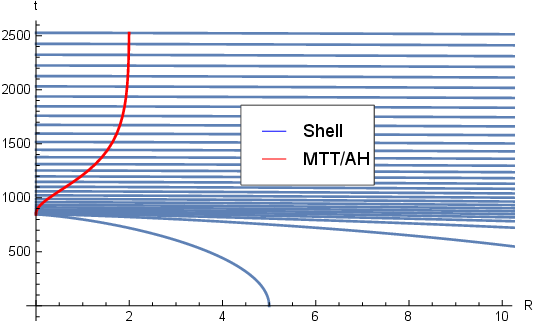}
\caption{}
\end{subfigure}
\begin{subfigure}{.45\textwidth}
\centering
\includegraphics[width=\linewidth]{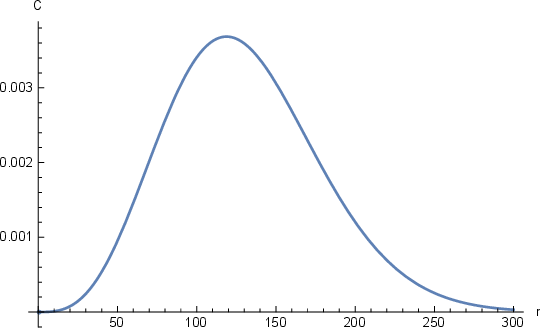}
\caption{}
\end{subfigure}
\caption{The Figures (a) shows the time evolution of the collapsing shells Eq. \eqref{tbc} and MTT Eq \eqref{t2mbc}, and (b) shows the dynamics of $C$ (or nature of MTT) for the Gaussian density distribution (\ref{margdensity1}) for bound collapse in $6$- dimensions. Figures (a) \& (b) shows that the MTT evolve as dynamical horizon, and  and remains positive throughout the collapse, which implies that the nature of MTT is spacelike and it becomes null (isolated horizon) as the matter stops falling in. 
}
\label{figure_bounded1}
\end{figure}
\begin{figure}[htb!]
\begin{subfigure}{.33\textwidth}
\centering
\includegraphics[width=\linewidth]{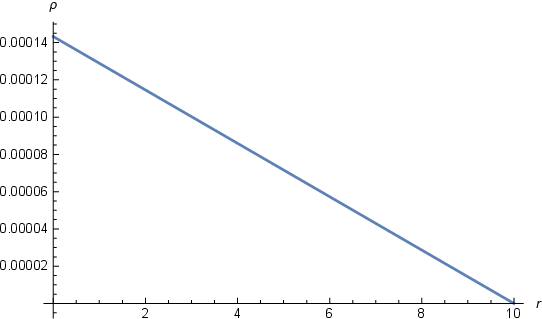}
\caption{}
\end{subfigure}
\begin{subfigure}{.33\textwidth}
\centering
\includegraphics[width=\linewidth]{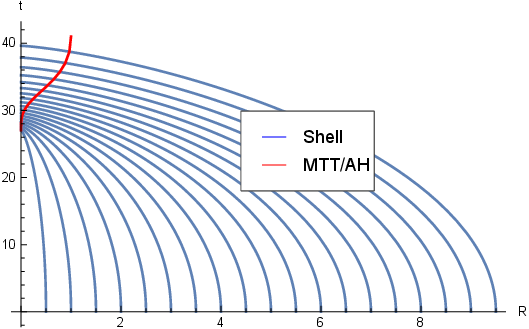}
\caption{}
\end{subfigure}
\begin{subfigure}{.31\textwidth}
\centering
\includegraphics[width=\linewidth]{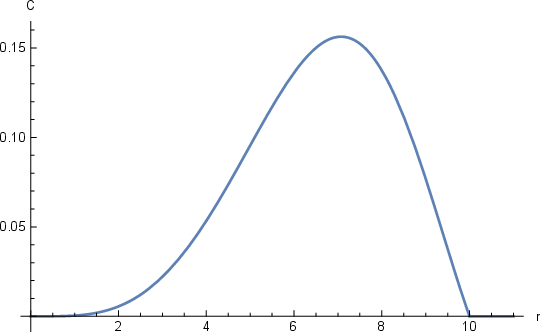}
\caption{}
\end{subfigure}
\caption{The figure (a) shows the density distribution $\rho_{1}$, Eq. \eqref{boundeddensity2} vs $r$. The Figures (b) shows the time evolution of the collapsing shells Eq. \eqref{tbc} and MTT Eq \eqref{t2mbc}, and (c) shows the dynamics of $C$ or MTT for the bound collapse in $6$- dimensions. Figs. (b) \& (c) shows that the MTT evolves as dynamical horizon and 
becomes null (isolated horizon) as the matter stops falling in. Here, we have chosen $m_{0}=1.0$.
}
\label{figure_bounded2}
\end{figure}
\begin{figure}[htb]
\begin{subfigure}{.33\textwidth}
\centering
\includegraphics[width=\linewidth]{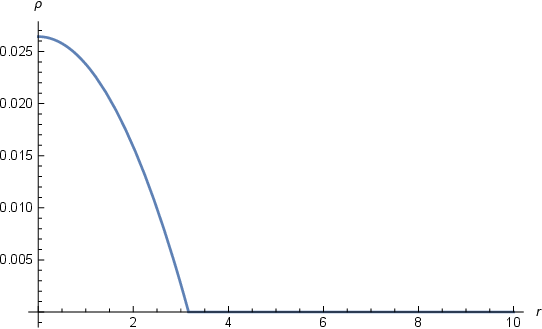}
\caption{}
\end{subfigure}
\begin{subfigure}{.33\textwidth}
\centering
\includegraphics[width=\linewidth]{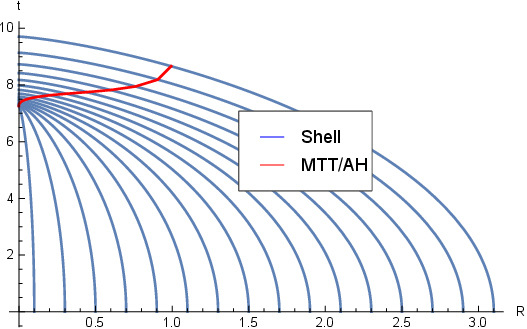}
\caption{}
\end{subfigure}
\begin{subfigure}{.31\textwidth}
\centering
\includegraphics[width=\linewidth]{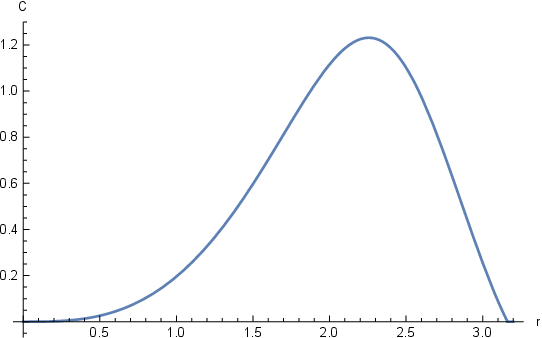}
\caption{}
\end{subfigure}
\caption{The figure (a) shows the density distribution $\rho_{2}$, Eq. \eqref{boundeddensity3} vs $r$. The Figures (b) shows the time evolution of the collapsing shells Eq. \eqref{tbc} and MTT Eq \eqref{t2mbc}, and (c) shows the dynamics of $C$ or MTT for the bound collapse in $6$- dimensions. Figs. (b) \& (c) shows that the MTT evolves as dynamical horizon and becomes null (isolated horizon) as the matter stops falling in. Here, we have chosen $m_{0}=1.0$.
}
\label{figure_bounded3}
\end{figure}
\section{Gravitational collapse of viscous matter}
For mathematical simplicity we will assume that the radial pressure is fixed $p_{r}=\frac{2(n-1)\eta\sigma}{(n-1)}+\zeta\theta$, whereas we shall continue with the non-zero combination $[p_{t}+\frac{2\eta\sigma}{(n-1)}-\zeta\theta]$ to keep the model physically viable. The components of the field equations are of the following form:
\begin{eqnarray}
    F'(t,r)&=&\frac{2\rho(t,r)R^{n-2}R'}{(n-2)(n-3)(n-4)}\,\, ,\\
    \dot{F}(t,r)&=& -\left[p_{r}(t,r)-\frac{2(n-2)\eta\sigma}{(n-1)}-\zeta\theta\right]\frac{2R^{n-2}\dot{R}}{(n-2)(n-3)(n-4)}=0\,\, ,\\
    \alpha'&=& \frac{(n-2)R'}{\rho R}\left[p_{t}+\frac{2\eta\sigma}{(n-1)}-\zeta\theta\right]\,\,,\\  2{\dot{R}}'&=&R'\frac{\dot{G}}{G}+\dot{R}\frac{H'}{H}\,\,,\\
    \label{massfunctionviscous1}F(t,r)&=&R^{n-5}(1-G+H)^{2}\,\, ,
\end{eqnarray}
Here, the number of unknown variables exceeds the independent field equations. The is constrained by using the equations of states $p_{t}=k_{t}\rho$, $p_{r}=k_{r}\rho$, and $\theta=k_{\theta}\rho$. Using these, the metric variables are of the following form,
\begin{eqnarray}\label{metricvariableviscous1}
    e^{2\alpha}=R^{2(n-2)a_{1}}, \hspace{2cm} e^{2\beta}=\frac{{R'}^{2}}{b(r)R^{2(n-2)a_{1}}},
\end{eqnarray}
where $a_{1}=k_{t}+2\eta k_{\sigma}/(n-1)-\zeta k_{\theta}$. Using (\ref{metricvariableviscous1}), the metric (\ref{metric}) becomes,
\begin{equation}
    ds^2=-R^{2(n-2)a_{1}}dt^2+\frac{{R'}^{2}}{b(r)R^{2(n-2)a_{1}}}dr^2+R^2(t,r)d\Omega^2_{n-2}\,\,.
\end{equation}
The equation of motion (\ref{massfunctionviscous1}) becomes,
\begin{equation}\label{eomviscous1}
    \dot{R}=-R^{(n-2)a_{1}}\left[\sqrt{\frac{F}{R^{n-5}}}+b(r)R^{2(n-2)a_{1}}-1\right]^{1/2}\,\,.
\end{equation}
For further simplification we choose the parameter $a_{1}=-(n-5)/2(n-2)$. Again the parametric choice for $R(t,r)$ is,
\begin{equation}
     R=\left[\frac{F}{b^{2}}\cos^{4}\left(\eta/2\right)\right]^{1/(n-5)}\,\,
\end{equation}
Using the above relations, the equation of collapsing shell (\ref{eomviscous1}) becomes,
\begin{equation}
    dt=\frac{2/(n-5)\left(F/b^{2}\right)^{(n-4)/(n-5)}\sin(\eta/2)\left(\cos(\eta/2)\right)^{(3n-11)/(n-5)}}{\left[(F/b)\cos^{2}(\eta/2)+b-(F/b^{2})\cos^{4}(\eta/2)\right]^{1/2}}d\eta\,\,.
\end{equation}
The solution of the above equation is given by,
\begin{equation}\label{tcv}
    t=-B\left(b\,R^{n-4}\,E-F^{(n-4)/(n-5)}b^{-(n-3)/(n-4)}J\right)\,\,,
\end{equation}
where $B=1/((n-4)\sqrt{b^{3}-F})$, and
\begin{eqnarray*}
    E&=&\, _2F_1\left[\frac{1}{2},\frac{2 (n-4)}{n-5};\frac{3 n-13}{n-5};-\frac{b F }{b^3-F}\left(\frac{b^2 R^{n-5}}{F}\right)^{1/2}\right]\\
    J&=&\, _2F_1\left[\frac{1}{2},\frac{2 (n-4)}{n-5};\frac{3 n-13}{n-5};-\frac{b F}{b^3-F}\right]\,,
\end{eqnarray*}
Again, above equation gives the decrease of radius $R(r,t)$ of the $r$th collapsing shell with time. 

\subsection*{Example}
The following mass profiles are taken as the initial density and the graphs below show how the matter collapse leads to formation of spacetime singularity and the local horizon (MTT).
\begin{eqnarray}
    \label{timeInddensity2}\rho_{1}(r)&=&\mathcal{N}\,m_{0}\left(1/r_{0}\right)^{n-1} e^{-r/r_{0}}\,\,,
\end{eqnarray}
where $\mathcal{N}=\frac{ (n-4) (n-3) (n-2) }{16 \pi  \Gamma (n-1)}$ is the normalisation factor.
\begin{figure}[htb]
\begin{subfigure}{.5\textwidth}
\centering
\includegraphics[width=\linewidth]{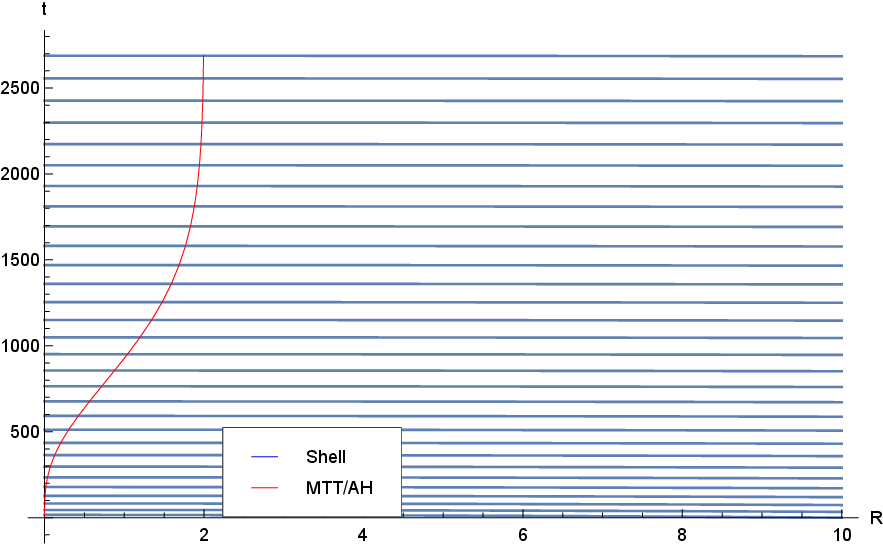}
\caption{}
\end{subfigure}
\begin{subfigure}{.45\textwidth}
\centering
\includegraphics[width=\linewidth]{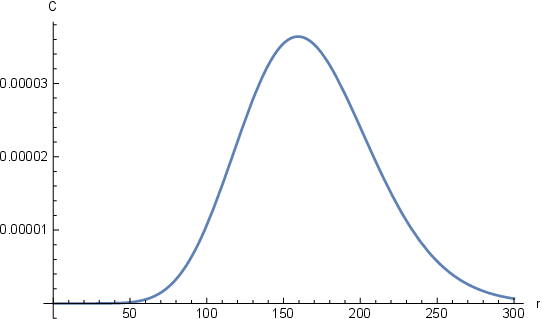}
\caption{}
\end{subfigure}
\caption{The Figures (a) shows the time evolution of the collapsing shells Eq. \eqref{tcv} and MTT, and (b) shows the dynamics of $C$ (or nature of MTT) for the Gaussian density distribution (\ref{margdensity1}) for viscous matter collapse in $6$- dimensions. Figures (a) \& (b) shows that the MTT evolve as dynamical horizon, and becomes null (isolated horizon) as the matter stops falling in. 
}
\label{figure_timeInd1}
\end{figure}
\begin{figure}[htb]
\begin{subfigure}{.33\textwidth}
\centering
\includegraphics[width=\linewidth]{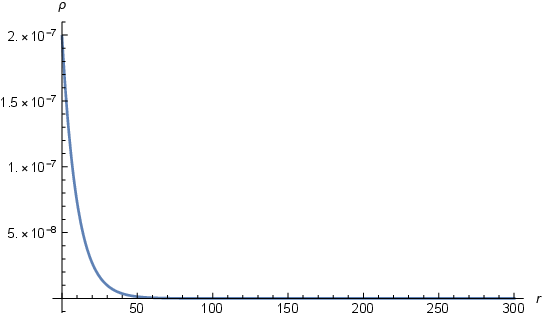}
\caption{}
\end{subfigure}
\begin{subfigure}{.33\textwidth}
\centering
\includegraphics[width=\linewidth]{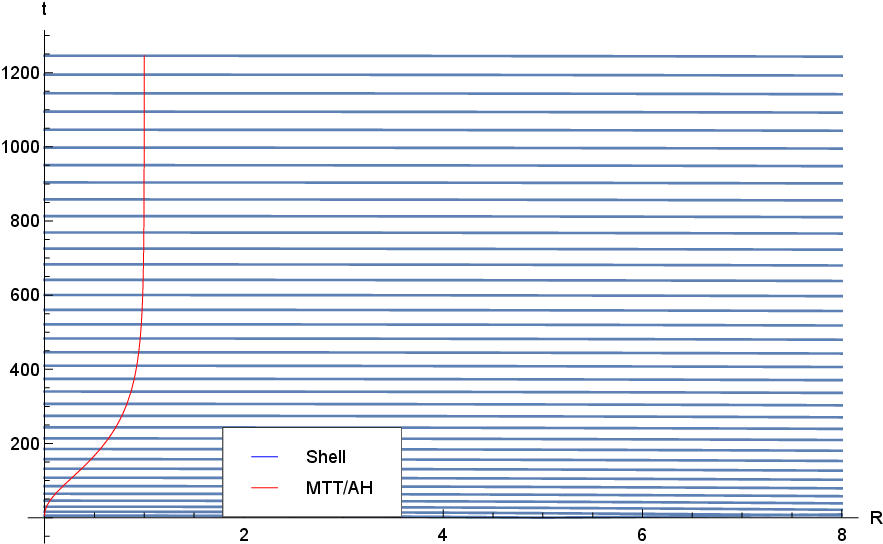}
\caption{}
\end{subfigure}
\begin{subfigure}{.31\textwidth}
\centering
\includegraphics[width=\linewidth]{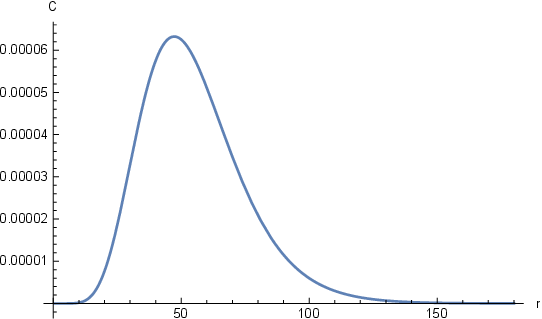}
\caption{}
\end{subfigure}
\caption{
The figure (a) shows the density distribution $\rho_{2}$, Eq. \eqref{timeInddensity2} vs $r$. The Figures (b) shows the time evolution of the collapsing shells Eq. \eqref{tcv} and MTT, and (c) shows the dynamics of $C$ or MTT for the viscous matter collapse in $6$- dimensions. Figs. (b) \& (c) shows that the MTT remains positive through collapse. This implies that MTT begins to evolve as dynamical horizon and becomes null (isolated horizon) as the matter stops falling in. Here, we have chosen $m_{0} = 1.0$, $r_{0} = 10.0\,m_{0}$.
}
\label{figure_timeInd2}
\end{figure}

\subsubsection*{A different choice: time dependent mass function}
One may approach the problem differently by generalising it to the full extent. The field equations of the following form, when system is in full generality, 
\begin{eqnarray}
    \rho(t,r)&=&\frac{(n-2)(n-3)(n-4)}{2}\frac{F'(t,r)}{R^{n-2}R'}\,\, ,\\
    p_{r}(t,r)&=&\frac{2(n-2)\eta\sigma}{(n-1)}+\zeta\theta-\frac{(n-2)(n-3)(n-4)}{2}\frac{\dot{F}(t,r)}{R^{n-2}\dot{R}}\,\, ,\\
    \alpha'&=& -\frac{(n-2)R'}{R}\frac{\left[p_{r}-p_{t}+2\eta\sigma \right]}{\left[\rho+p_{r}-\frac{2(n-2)\eta\sigma}{n-1}-\zeta\theta \right]}-\frac{\left[p_{r}-\frac{2(n-2)\eta\sigma}{n-1}-\zeta\theta \right]'}{\left[\rho+p_{r}-\frac{2(n-2)\eta\sigma}{n-1}-\zeta\theta \right]}  \,\,,\\
    2{\dot{R}}'&=&R'\frac{\dot{G}}{G}+\dot{R}\frac{H'}{H}\,\,,\\
    \label{massfunctionviscous2}F(t,r)&=&R^{n-5}(1-G+H)^{2}\,\, ,
\end{eqnarray} 
To solve these equations, we assume the set of equation of states, $p_{t}=k_{t}\rho$, $p_{r}=k_{r}\rho$ and $\theta=k_{\theta}\rho$. Using these, the metric functions are of following form, 
\begin{eqnarray}\label{metricvariableviscous2}
    e^{2\alpha}=\frac{R^{2(n-2)a_{1}}}{\rho^{2a_{2}}}, \hspace{2cm} e^{2\beta}=\frac{R'^{2}}{1+r^{2}B(t,r)}
\end{eqnarray}
where $a_{1}$ and $a_{2}$ defined as,
\begin{eqnarray}
    a_{1}=\frac{k_{t}-k_{r}+2\eta k_{\sigma}}{1+k_{r}-2(n-2)\eta k_{\sigma}/(n-1)-\zeta k_{\theta}}, \hspace{2cm} a_{2}=\frac{k_{r}-2(n-2)\eta k_{\sigma}/(n-1)-\zeta k_{\theta}}{1+k_{r}-2(n-2)\eta k_{\sigma}/(n-1)-\zeta k_{\theta}}
\end{eqnarray}
The metric (\ref{metric}) for this space time becomes,
\begin{equation}
    ds^2=-\frac{R^{2(n-2)a_{1}}}{\rho^{2a_{2}}}dt^2+\frac{R'^{2}}{1+r^{2}B(t,r)}dr^2+R^2(t,r)d\Omega^2_{n-2}\,\,.
\end{equation}
The equation of motion (\ref{massfunctionviscous2}) becomes,
\begin{equation}\label{eomviscous2}
   \dot{R}=-R^{2(n-2)a_{1}}\rho^{-2a_{2}}\left[\sqrt{\frac{F}{R^{n-5}}}+r^{2}B(t,r)\right]^{1/2}\,\,. 
\end{equation}
Let us assume the functions in above equation are of separable type,
\begin{eqnarray}
      F(t,r)=F_{1}(r)F_{2}(t),\hspace{0.75cm}B(t,r)=B_{1}(r)B_{2}(t), \hspace{0.75cm}\rho(t,r)=\rho_{1}(r)\rho_{2}(t), 
\end{eqnarray}
also we can define,
\begin{eqnarray}
     B_{1}(r)=k(r)/r^{2}, \hspace{2cm}B_{2}(t)=-\sqrt{F_{2}(t)}=-\rho_{2}(t)^{2a_{2}}\,\,.
\end{eqnarray}
The parametric choice for $R(t,r)$ is given by,
\begin{equation}
    R=\left[\frac{F_{1}}{k^{2}}\cos^{4}\left(\eta/2 \right)\right]^{1/(n-5)}\,\,,    
\end{equation}
Using above relations, the equation for collapsing shell (\ref{eomviscous2}) becomes,
\begin{equation}
    dt=\frac{2{\rho_{1}}^{a_{2}}}{(n-5)\sqrt{k}}\left[\frac{F_{1}}{k}\cos^{4}(\eta/2)\right]^{(1-(n-2)a_{1})/(n-5)}\,\,d\eta\,\,.
\end{equation}
The solution of above equation is,
\begin{eqnarray}\label{tcan}
  t&=&-Q\left(k^{S/(n-5)}T-\sqrt{\pi}k^{U/2(n-5)}V\right)\,\,,
\end{eqnarray}
where,
\begin{eqnarray*}
    Q=(4\rho^{a_{2}}/(5(n-5)-4a_{1}(n-2)))F_{1}^{(1-a_{1}(n-2))/(n-5)}, ~~~~S=(2n-11-a_{1}(n-2)),~~~
    T=(-n+3+2a_{1}(n-2)),
\end{eqnarray*}
and
\begin{eqnarray*}
    U&=&\, _2F_1\left[\frac{1}{2},\frac{5}{2}-\frac{2 a_{1} (n-2)}{n-5};\frac{7}{2}-\frac{2 a_{1} (n-2)}{n-5};\left(\frac{k^2 R^{n-5}}{F}\right)^{1/2}\right]\,,\\
    V&=&\frac{\Gamma \left[\frac{4 n a_{1}}{10-2 n}-\frac{8 a_{1}}{10-2 n}-\frac{7 n}{10-2 n}+\frac{35}{10-2 n}\right]}{\Gamma \left[-\frac{2 n a_{1}}{n-5}+\frac{4 a_{1}}{n-5}+\frac{3 n}{n-5}-\frac{15}{n-5}\right]}\,\,.
\end{eqnarray*}
\subsection*{Example}
Let us demonstrate the above discussion with the following example of initial density. The graphs show how the matter collapse leads to a 
spacetime singularity and a horizon.

\begin{figure}[htb]
\begin{subfigure}{.5\textwidth}
\centering
\includegraphics[width=\linewidth]{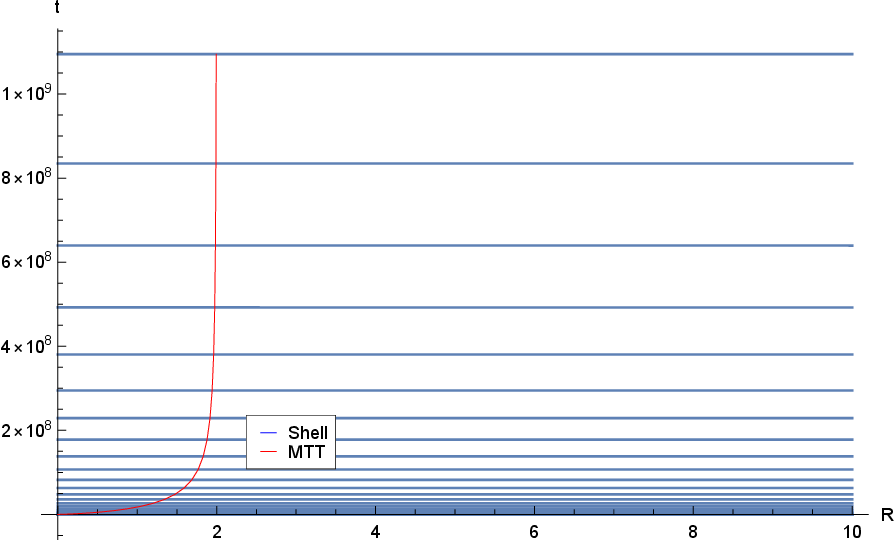}
\caption{}
\end{subfigure}
\begin{subfigure}{.45\textwidth}
\centering
\includegraphics[width=\linewidth]{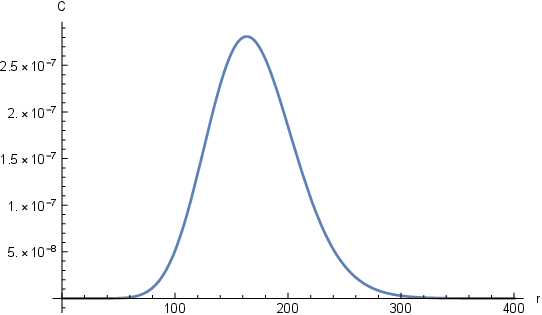}
\caption{}
\end{subfigure}
\caption{The Figures (a) shows the time evolution of the collapsing shells Eq. \eqref{tcan} and MTT, and (b) shows the dynamics of $C$ (or nature of MTT) for the Gaussian density distribution (\ref{margdensity1}) for anisotropic collapse in $6$- dimensions. Figure (b) shows that the MTT remains positive throughout the collapse and becomes null as matter stops to fall in. Figure (a) shows that as the matter starts to fall in, MTT begins to evolve as dynamical horizon, and becomes null isolated horizon as the matter stops falling in. Here, we have chosen $k_{t}=1/4$, $k_{r}=1/2$, $k_{\sigma}=1/4$, $k_{\theta}=3/2$, $\zeta=1/2$, $\eta=1/16$ and $\chi=1$.}
\label{figure_timeDpn1}
\end{figure}
\subsection*{A simpler choice: perfect fluids}
The perfect fluid case arise when pressure components are equal i.e. $p_{t}=p_{r}=p$ and viscosity parameters are zero i.e. $\eta=\zeta=0$. The field equations (\ref{H00})-(\ref{mass function}) reduce to,
    \begin{eqnarray}
        \rho(t,r)&=&\frac{(n-2)(n-3)(n-4)}{2}\frac{F'(t,r)}{R^{n-2}R'}\,\, ,\\
        p(t,r)&=&-\frac{(n-2)(n-3)(n-4)}{2}\frac{\dot{F}(t,r)}{R^{n-2}\dot{R}}\,\, ,\\
        \alpha'&=& -\frac{p'}{\rho+p}  \,\,,\\
        2{\dot{R}}'&=&R'\frac{\dot{G}}{G}+\dot{R}\frac{H'}{H}\,\,,\\
        F(t,r)&=&R^{n-5}(1-G+H)^{2}\,\, ,
    \end{eqnarray}
    Above set of equation can be solved by using equation of state $p=k_{p}\rho$, using this the metric functions can be written in the following form,
\begin{eqnarray}
    e^{2\alpha}=\rho^{-2a_{1}}\,\,, \hspace{2cm} e^{2\beta}=\frac{{R'}^{2}}{1+r^{2}B(t,r)}\,\,,
\end{eqnarray}
where $a_{1}=k_{p}/(1+k_{p})$. The metric (\ref{metric}) becomes,
\begin{equation}
    ds^2=-\frac{1}{\rho^{2a_{1}}}dt^2+\frac{{R'}^{2}}{1+r^{2}B(t,r)}dr^2+R^2(t,r)d\Omega^2_{n-2}\,\,.
\end{equation}
The equation of motion (\ref{eom}) becomes,
\begin{equation}\label{eomperfectfluid}
    \dot{R}=-\rho^{-a_{1}}\left[\sqrt{\frac{F}{R^{n-5}}}+r^{2}B(t,r)\right]^{1/2}\,\,.
\end{equation}
We can take separable type variables for the simplification of the above equation,
\begin{eqnarray}
      F(t,r)=F_{1}(r)F_{2}(t),\hspace{0.75cm}B(t,r)=B_{1}(r)B_{2}(t), \hspace{0.75cm}\rho(t,r)=\rho_{1}(r)\rho_{2}(t), 
\end{eqnarray}
also, we can define,
\begin{eqnarray}
     B_{1}(r)=k(r)/r^{2}, \hspace{2cm}B_{2}(t)=-\sqrt{F_{2}(t)}=-\rho_{2}(t)^{2a_{2}}
\end{eqnarray}
again parametric choice for $R(t,r)$,
\begin{equation}
    R=\left[\frac{F_{1}}{k^{2}}\cos^{4}\left(\eta/2\right)\right]^{1/(n-5)}\,\,,    
\end{equation}
The equation of motion (\ref{eomperfectfluid}) becomes,
\begin{equation}
    dt=\frac{2{\rho_{1}}^{a_{1}}}{n-5}\left[F_{1}/k^{(n-4)/2}\right]^{1/(n-5)}\left[\cos^{4}(\eta/2)\right]^{1/(n-5)}d\eta\,\,,
\end{equation}
Solution for above equation is following:
\begin{eqnarray}\label{tcp}
  t&=&W \left[\mathcal{F}(r)-\mathcal{F}(R) \right]\nonumber\\
\end{eqnarray}
where $W=(4{\rho_{1}}^{a_{1}}/(n-1))F_{1}^{-1/4}k^{3/2(n-5)}$, and
\begin{equation}
   \mathcal{F}(x)= x^{(n-1)/4}\, ~~{}_2F_1\left[\frac{1}{2},\frac{1}{2}+\frac{2}{n-5};\frac{3}{2}+\frac{2}{n-5};\left(\frac{k^{2} x^{n-5}}{F_{1}}\right)^{1/2}\right],
\end{equation}
 The above solution gives the motion of collapsing shells, and we can get the equation for horizon (MTT) by using equation(\ref{mtt}).
\subsection*{Example}
Again, as in the above examples, we discuss the collapse of matter shells 
by taking the initial matter density in the following form:
\begin{figure}[htb]
\begin{subfigure}{.5\textwidth}
\centering
\includegraphics[width=\linewidth]{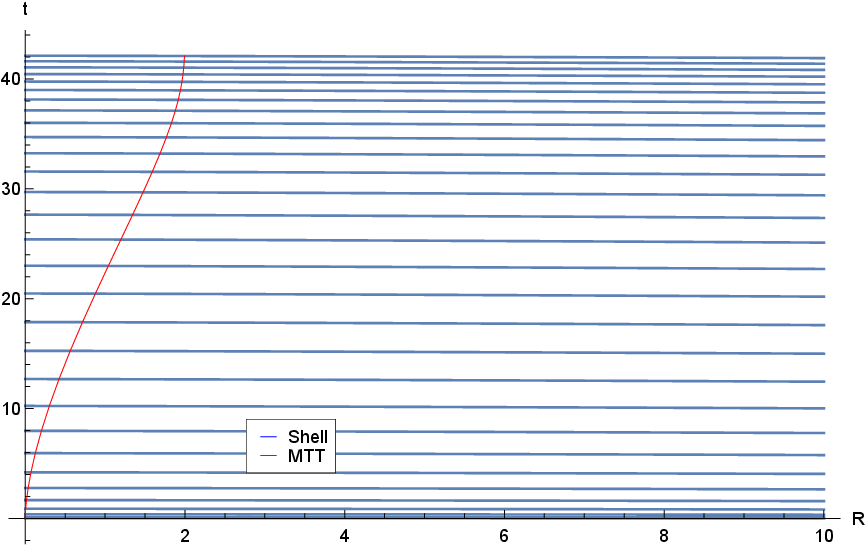}
\caption{}
\end{subfigure}
\begin{subfigure}{.45\textwidth}
\centering
\includegraphics[width=\linewidth]{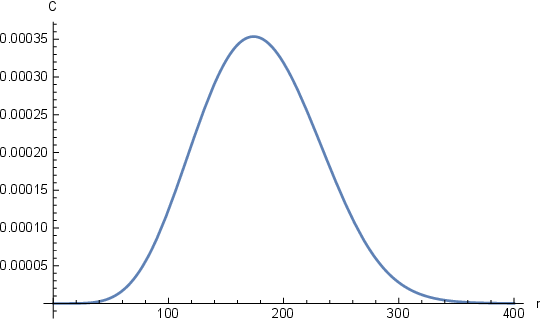}
\caption{}
\end{subfigure}
\caption{The Figures (a) shows the time evolution of the collapsing shells Eq. \eqref{tcp} and MTT, and (b) shows the dynamics (nature) of $C$ (MTT) for the Gaussian density distribution \eqref{margdensity1} for perfect collapse in $6$- dimensions. Figure (b) shows that the MTT remains positive throughout the collapse and becomes null as matter stops to fall in. Figure (a) shows that as the matter starts to fall in, MTT begins to evolve as dynamical horizon, and becomes null isolated horizon as the matter stops falling in. Here, we have chosen $k_{p}=1/100$ and $\chi=1$.
}
\label{figure_perfectfluid1}
\end{figure}
\begin{figure}[htb]
\begin{subfigure}{.5\textwidth}
\centering
\includegraphics[width=\linewidth]{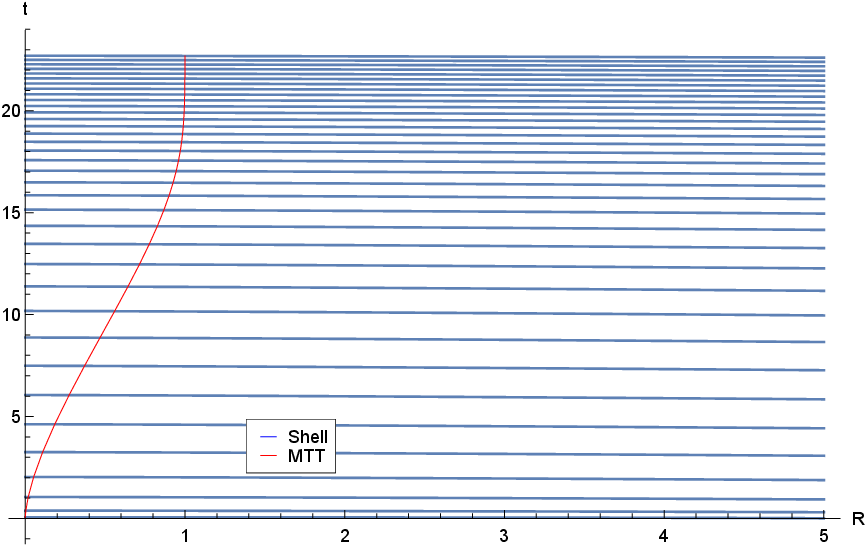}
\caption{}
\end{subfigure}
\begin{subfigure}{.45\textwidth}
\centering
\includegraphics[width=\linewidth]{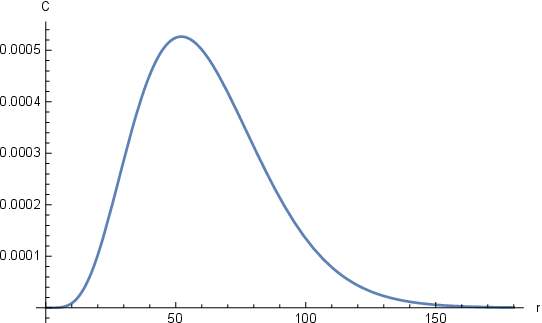}
\caption{}
\end{subfigure}
\caption{The Figures (a) shows the time evolution of the collapsing shells Eq. \eqref{tcp} and MTT, and (b) shows the dynamics (nature) of $C$ (MTT) for the density distribution Eq. \eqref{timeInddensity2} for perfect collapse in $6$- dimensions. Figure (b) shows that the MTT remains positive throughout the collapse and becomes null as matter stops to fall in. Figure (a) shows that as the matter starts to fall in, MTT begins to evolve as dynamical horizon, and becomes null isolated horizon as the matter stops falling in.
}
\label{figure_perfectfluid2}
\end{figure}

\section{Discussion}
In the study carried out above for the gravitational collapse of matter fields in the GB theory, we have determined the formation of the central singularity, along with the time of formation of the black hole horizon. Note that the time of formation of the horizon and the singularity time increases if the fluid is viscous (as compared to the dust models). This is expected since viscosity is well known to hamper the collapse rates of fluids. The studies also indicate the rate of growth of the black hole horizon as matter falls in the horizon. The equations and the graphical depiction in the previous sections should convince the reader that in the GB theory too, the censorship conjecture holds true. While it may correct that some peculiar matter profiles may be conjured up where the singularity may be visible, the smooth and physically viable profiles considered here indicates the robustness of the censorship conjecture in higher dimensions and beyond GR. 
\section{Appendix}
\subsection{Field equations}
The field equations for the pure GB theory are quite complicated. In this section, we have collected the non-zero components of the GB tensor
for the benefit of the reader.
\begin{eqnarray}                 
H_{ab}&=&\frac{2(n-3)(n-4)}{R}\bigg[R{\dot{R}}^{2}\ddot{\beta}e^{-4\alpha}- 
           R{R'}^{2}\alpha''e^{-4\beta} + (n-5){R'}^{2}\ddot{R}e^{-2\beta-2\alpha} -2R{\dot{R'}}^{2}e^{-2\alpha-2\beta}\nonumber\\
          &&+(n-5){\dot{R}}^{2}R''e^{-2\beta-2\alpha}-R\ddot{\beta}e^{-2\alpha}+R\alpha''e^{-2\beta}-R\alpha'\beta'e^{-2\beta}+R\dot{\beta}\dot{\alpha}e^{-2\alpha}-R{\dot{\beta}}^{2}{\dot{R}}^{2}e^{-4\alpha}\nonumber\\
          &&-R{\alpha'}^{2}{R'}^{2}e^{-4\beta}-2R\dot{\beta}\dot{R}\ddot{R}e^{-4\alpha}+R{R'}^{2}\ddot{\beta}e^{-2\alpha-2\beta}+R{\dot{R}}^{2}\alpha''e^{-2\alpha-2\beta}-2R\alpha'R'R''e^{-4\beta}+2RR''\ddot{R}e^{-2\alpha-2\beta}\nonumber\\
          &&-R\dot{\beta}\dot{\alpha}{R'}^{2}e^{-2\alpha-2\beta}-R\beta'\alpha'{\dot{R}}^{2}e^{-2\alpha-2\beta}+2R\beta'R'\dot{\alpha}\dot{R}e^{-2\alpha-2\beta}-2R\alpha'R'\dot{\beta}\dot{R}e^{-2\alpha-2\beta}-(n-5)\dot{\beta}{\dot{R}}^{3}e^{-4\alpha}\nonumber\\
          &&+(n-5)\dot{\alpha}{\dot{R}}^{3}e^{-4\alpha}+(n-5)\beta'{R'}^{3}e^{-4\beta}-(n-5)\alpha'{R'}^{3}+3R\dot{\alpha}\dot{\beta}{\dot{R}}^{2}e^{-4\alpha}-R{\dot{\beta}}^{2}{R'}^{2}e^{-2\alpha-2\beta}\nonumber\\
          &&-R{\alpha'}^{2}{\dot{R}}^{2}e^{-2\alpha-2\beta}+3R\alpha'\beta'{R'}^{2}e^{-4\beta}-(n-5)\dot{\alpha}\dot{R}{R'}^{2}+(n-5)\dot{\beta}\dot{R}{R'}^{2}e^{-2\alpha-2\beta}-(n-5)\beta'R'{\dot{R}}^{2}e^{-2\alpha-2\beta}\nonumber\\
          &&+(n-5)\alpha'R'{\dot{R}}^{2}e^{-2\alpha-2\beta}-2R\beta'R'\ddot{R}e^{-2\alpha-2\beta}+4R\dot{\beta}R'{\dot{R}}'e^{-2\alpha-2\beta}-2R\dot{\alpha}\dot{R}R''e^{-2\alpha-2\beta}+4R\alpha'\dot{R}{\dot{R}}'e^{-2\alpha-2\beta}\nonumber\\
          &&+R{\alpha'}^{2}e^{-2\beta}-R{\dot{\beta}}^{2}e^{-2\alpha}-(n-5)\dot{\beta}\dot{R}e^{-2\alpha}+(n-5)\dot{\alpha}\dot{R}e^{-2\alpha}+(n-5)\alpha'R'e^{-2\beta}-(n-5)\beta'R'e^{-2\beta}\nonumber\\
          &&-(n-5)\ddot{R}e^{-2\alpha}+(n-5)R''e^{-2\beta}-(n-5){\dot{R}}^{2}\ddot{R}e^{-4\alpha}-(n-5){R'}^{2}R''e^{-4\beta}-\frac{(n-5)(n-6)}{4R}{R'}^{4}e^{-4\beta}\nonumber\\
          &&-\frac{(n-5)(n-6)}{4R}{\dot{R}}^{4}e^{-4\alpha}-\frac{(n-5)(n-6)}{4R}+\frac{(n-5)(n-6)}{2R}{R'}^{2}e^{-2\beta}-\frac{(n-5)(n-6)}{2R}{\dot{R}}^{2}e^{-2\alpha}\nonumber\\
          &&+\frac{(n-5)(n-6)}{2R}{\dot{R}}^{2}{R'}^{2}\bigg] \,\,,\\
H_{00}&=&\frac{2e^{-4\beta-2\alpha}(n-2)(n-3)(n-4)}{R^{4}}\bigg[-   
          R{R'}^{3}\beta'e^{4\alpha}-R\dot{R}{R'}^{2}\dot{\beta}e^{2\beta+2\alpha}+R{\dot{R}}^{2}\beta'R'e^{2\beta+2\alpha}+R{\dot{R}}^{3}\dot{\beta}e^{4\beta}\nonumber\\
          &&+\frac{(n-5)}{4}{R'}^{4}e^{4\alpha}+R{R'}^{2}R''e^{4\alpha}-\frac{(n-5)}{2}{\dot{R}}^{2}{R'}^{2}e^{2\beta+2\alpha}-R{\dot{R}}^{2}R''e^{2\beta+2\alpha}+\frac{(n-5)}{4}{\dot{R}}^{4}e^{4\beta}\nonumber\\
       &&+R\beta'R'e^{2\beta+4\alpha}+R\dot{\beta}\dot{R}e^{4\beta+2\alpha}-\frac{(n-5)}{2}{R'}^{2}-RR''e^{2\beta+4\alpha}+\frac{(n-5)}{2}{\dot{R}}^{2}e^{4\beta+2\alpha}\nonumber\\
          &&+\frac{(n-5)}{4}e^{4\alpha+4\beta}\bigg]\,\,,
          \end{eqnarray}
   \begin{eqnarray}
H_{11}&=&\frac{2(n-2)(n-3)(n-4)}{R^{4}}\bigg[-R{R'}^{3}\alpha'e^{-2\beta} + 
          R{\dot{R}}^{2}R'\alpha'e^{-2\alpha} + R{\dot{R}}^{3}\dot{\alpha}e^{-4\alpha+2\beta}-R\dot{R}\dot{\alpha}{R'}^{2}e^{-2\alpha}\nonumber\\
          &&-\frac{(n-5)}{4}{R'}^{4}e^{-2\beta}-\frac{(n-5)}{4}{\dot{R}}^{4}e^{-4\alpha+2\beta}-R{\dot{R}}^{2}\ddot{R}e^{-4\alpha+2\beta}+\frac{(n-5)}{4}{\dot{R}}^{2}{R'}^{2}e^{-2\alpha}+R{R'}^{2}\ddot{R}e^{-2\alpha}\nonumber\\
          &&+RR'\alpha'+R\dot{\alpha}\dot{R}e^{2\beta-2\alpha}-\frac{(n-5)}{2}{\dot{R}}^{2}e^{2\beta-2\alpha}-R\ddot{R}e^{2\beta-2\alpha}+\frac{(n-5)}{2}{R'}^{2}+\frac{(n-5)}{4}e^{2\beta}\bigg]\,\,,\\
H_{01}&=&H_{10}=\frac{2(n-2)(n-3)(n-4)}{R^{3}}\left[\alpha'\dot{R} + 
          \dot{\beta} R'-{\dot{R}}'\right]\left[1+{\dot{R}}^{2}e^{-2\alpha}-{R'}^{2}e^{-2\beta}\right] \,\,,
            \end{eqnarray}

\subsection{Calculating the value of C }
The nature of the MTT is determined by the signature of $C$: the horizon is null if $C=0$, and spacelike if $C>0$. Therefore, it is imperative that
the values of $C$ be determined to ascertain how the nature of matter fields control the growth of MTTs.

\begin{equation}\label{c}
    C=\frac{R_{\mu\nu}l^{\mu}l^{\nu}}{\mathcal{R}/2-\left(R_{s}/{2}+R_{\mu\nu}l^{\mu}n^{\nu}\right)}
\end{equation}
where $\mathcal{R}$ is Ricci scalar for $(n-2)$ sphere, $R_{s}$ is the Ricci scalar of full space time, $R_{\mu\nu}$ is Ricci tensor of full space time, $l^{\mu}$ and $n^{\mu}$ are the outgoing and ingoing null normals to $(n-2)$ sphere respectively. By substituting these values and Ricci scalar $\mathcal{R}=[(n-2)(n-3)]/r^{2}$, where $r$ is the radius of the $(n-2)$ sphere.
For the pressureless collapse the expression for the null normals are,
\begin{eqnarray}
    l^{\mu}&=&\delta_{t}^{\mu}+e^{-\beta(t,r)}\delta_{r}^{\mu}\\
    n^{\mu}&=&\frac{1}{2}\delta_{t}^{\mu}-\frac{1}{2}e^{-\beta(t,r)}\delta_{r}^{\mu}
\end{eqnarray}
and 
\begin{eqnarray}
    R_{\mu\nu}l^{\mu}l^{\nu}&=&\frac{\rho R_{2m}^{2}}{2(n-3)(n-4)}\left(\frac{R_{2m}^{(n-5)}}{F}\right)^{1/2}\\
    R_{\mu\nu}l^{\mu}n^{\nu}&=&\frac{(n-1)(n-5)}{8R_{2m}^{2}}\left(\frac{F}{R_{2m}^{(n-5)}}\right)^{1/2}-\frac{3}{4}\frac{\rho R_{2m}^{2}}{(n-2)(n-3)(n-4)}\left(\frac{R_{2m}^{(n-5)}}{F}\right)^{1/2}
\end{eqnarray}
now substituting above expressions into equation (\ref{c}),
\begin{equation}
    C=\frac{2F'R_{2m}^{4}}{4(n-3)R'R_{2m}^{(n+1)/2} F^{1/2}-F'R_{2m}^{4}-2(n-1)R'FR_{2m}^{3}}
\end{equation}
For viscous fluid when system is in full generality the null normals may be taken to have the following form:
\begin{eqnarray}
    l^{\mu}&=&\frac{1}{2\chi}\left(\delta_{t}^{\mu}+\delta_{r}^{\mu}\right)\\
    n^{\mu}&=&\chi\left(\delta_{t}^{\mu}+\delta_{r}^{\mu}\right).
\end{eqnarray}
This gives us the following expressions for the Riemann components:
\begin{eqnarray}
    R_{\mu\nu}l^{\mu}l^{\nu}&=&\frac{1}{4\chi^{2}}\frac{R_{2m}^{(n-1)/2}}{2(n-3)(n-4)F^{1/2}}\left[\rho+p_{r}-2(n-2)\eta\sigma/(n-1)-\zeta\theta\right]\\
    R_{\mu\nu}l^{\mu}n^{\nu}&=&\frac{\left[p_{t}+2\eta\sigma/(n-1)-\zeta\theta\right]}{2(n-3)(n-4)}\left(\frac{R_{2m}^{(n-5)}}{F}\right)^{1/2}-\frac{3 R_{2m}^{(n-1)/2}}{4(n-2)(n-3)(n-4)F^{1/2}}\left[\rho-p_{r}+2(n-2)\eta\sigma/(n-1)+\zeta\theta\right]\nonumber\\
    &&-\frac{\rho R_{2m}^{3(n-1)/2}}{2[(n-2)(n-3)(n-4)]^{2}F^{3/2}}\left[p_{r}-2(n-2)\eta\sigma/(n-1)-\zeta\theta\right]+\frac{(n-1)(n-5)F^{1/2}}{8R_{2m}^{(n-1)/2}}
\end{eqnarray}
Again using above expressions into equation (\ref{c}),
\begin{eqnarray}\label{Cgeneral}
    C=\frac{A_{1}\left[\rho+p_{r}-2(n-2)\eta\sigma/(n-1)-\zeta\theta\right]}{B_{1}-E_{1}\left[\rho-p_{r}+2(n-2)\eta\sigma/(n-1)+\zeta\theta\right]-J_{1}}
\end{eqnarray}
whereas for time independent case, the expression for $C$ is following,
\begin{equation}\label{Creduced}
    C=\frac{A_{1}\,\rho }{B_{1}-E_{1}\,\rho -J_{1}}\,\,,
\end{equation}
where,
\begin{eqnarray*}
    A_{1}&=&\frac{1}{4\chi^{2}}\frac{R_{2m}^{(n-1)/2}}{2(n-3)(n-4)F^{1/2}}, ~~~~B_{1}=\frac{(n-2)(n-3)}{2R_{2m}^{2}},~~~
    E_{1}=\frac{R_{2m}^{(n-1)/2}}{4(n-3)(n-4)F^{1/2}}, ~~~~
    J_{1}=\frac{(n-1)(n-2)F^{1/2}}{4R_{2m}^{(n-1)/2}}.
\end{eqnarray*}
\subsection{Expressions for curvature using matter variable}
From the above expressions for $C$ in Eq. \eqref{Cgeneral} and \eqref{Creduced}, we can see that the expressions are in terms of matter variables and mass function, but the original expression \eqref{c} was in terms of curvature terms. In the following, we rewrite the curvature terms using  matter variables.

\begin{equation}
    R_{0101}=-{\dot{\beta}}^{2}e^{2\beta}-\ddot{\beta}e^{2\beta}+\dot{\beta}\dot{\alpha}e^{2\beta}-\beta'\alpha'e^{2\alpha}+{\alpha'}^{2}e^{2\alpha}+{\alpha''}e^{2\alpha}
\end{equation}
Now again,
\begin{eqnarray}
          H_{22}&=&\frac{2(n-3)(n-4)}{R}\bigg[-R{\dot{R}}^{2}\ddot{\beta}e^{-4\alpha}-R{R'}^{2}\alpha''e^{-4\beta}+(n-5){R'}^{2}\ddot{R}e^{-2\beta-2\alpha}-2R{\dot{R'}}^{2}e^{-2\alpha-2\beta}\nonumber\\
          &&+(n-5){\dot{R}}^{2}R''e^{-2\beta-2\alpha}-R\ddot{\beta}e^{-2\alpha}+R\alpha''e^{-2\beta}-R\alpha'\beta'e^{-2\beta}+R\dot{\beta}\dot{\alpha}e^{-2\alpha}-R{\dot{\beta}}^{2}{\dot{R}}^{2}e^{-4\alpha}\nonumber\\
          &&-R{\alpha'}^{2}{R'}^{2}e^{-4\beta}-2R\dot{\beta}\dot{R}\ddot{R}e^{-4\alpha}+R{R'}^{2}\ddot{\beta}e^{-2\alpha-2\beta}+R{\dot{R}}^{2}\alpha''e^{-2\alpha-2\beta}-2R\alpha'R'R''e^{-4\beta}+2RR''\ddot{R}e^{-2\alpha-2\beta}\nonumber\\
          &&-R\dot{\beta}\dot{\alpha}{R'}^{2}e^{-2\alpha-2\beta}-R\beta'\alpha'{\dot{R}}^{2}e^{-2\alpha-2\beta}+2R\beta'R'\dot{\alpha}\dot{R}e^{-2\alpha-2\beta}-2R\alpha'R'\dot{\beta}\dot{R}e^{-2\alpha-2\beta}-(n-5)\dot{\beta}{\dot{R}}^{3}e^{-4\alpha}\nonumber\\
          &&+(n-5)\dot{\alpha}{\dot{R}}^{3}e^{-4\alpha}+(n-5)\beta'{R'}^{3}e^{-4\beta}-(n-5)\alpha'{R'}^{3}+3R\dot{\alpha}\dot{\beta}{\dot{R}}^{2}e^{-4\alpha}-R{\dot{\beta}}^{2}{R'}^{2}e^{-2\alpha-2\beta}\nonumber\\
          &&-R{\alpha'}^{2}{\dot{R}}^{2}e^{-2\alpha-2\beta}+3R\alpha'\beta'{R'}^{2}e^{-4\beta}-(n-5)\dot{\alpha}\dot{R}{R'}^{2}+(n-5)\dot{\beta}\dot{R}{R'}^{2}e^{-2\alpha-2\beta}-(n-5)\beta'R'{\dot{R}}^{2}e^{-2\alpha-2\beta}\nonumber\\
          &&+(n-5)\alpha'R'{\dot{R}}^{2}e^{-2\alpha-2\beta}-2R\beta'R'\ddot{R}e^{-2\alpha-2\beta}+4R\dot{\beta}R'{\dot{R}}'e^{-2\alpha-2\beta}-2R\dot{\alpha}\dot{R}R''e^{-2\alpha-2\beta}+4R\alpha'\dot{R}{\dot{R}}'e^{-2\alpha-2\beta}\nonumber\\
          &&+R{\alpha'}^{2}e^{-2\beta}-R{\dot{\beta}}^{2}e^{-2\alpha}-(n-5)\dot{\beta}\dot{R}e^{-2\alpha}+(n-5)\dot{\alpha}\dot{R}e^{-2\alpha}+(n-5)\alpha'R'e^{-2\beta}-(n-5)\beta'R'e^{-2\beta}\nonumber\\
          &&-(n-5)\ddot{R}e^{-2\alpha}+(n-5)R''e^{-2\beta}-(n-5){\dot{R}}^{2}\ddot{R}e^{-4\alpha}-(n-5){R'}^{2}R''e^{-4\beta}-\frac{(n-5)(n-6)}{4R}{R'}^{4}e^{-4\beta}\nonumber\\
          &&-\frac{(n-5)(n-6)}{4R}{\dot{R}}^{4}e^{-4\alpha}-\frac{(n-5)(n-6)}{4R}+\frac{(n-5)(n-6)}{2R}{R'}^{2}e^{-2\beta}-\frac{(n-5)(n-6)}{2R}{\dot{R}}^{2}e^{-2\alpha}\nonumber\\
          &&+\frac{(n-5)(n-6)}{2R}{\dot{R}}^{2}{R'}^{2}\bigg]=R^{2}\left[p_{t}+\frac{2}{(n-1)}\eta\sigma-\zeta\theta\right]
\end{eqnarray}
Using, $\dot{R'}=\dot{\beta}R'+\alpha'\dot{R}$ and equation (\ref{mass function})
\begin{eqnarray}
    \Rightarrow&& \sqrt{\frac{F}{R^{n-5}}}e^{-2\alpha-2\beta}\left[-{\dot{\beta}}^{2}e^{2\beta}-\ddot{\beta}e^{2\beta}+\dot{\beta}\dot{\alpha}e^{2\beta}-\beta'\alpha'e^{2\alpha}+{\alpha'}^{2}e^{2\alpha}+{\alpha''}e^{2\alpha}\right]-\frac{(n-5)(n-6)}{4R}\frac{F}{R^{n-5}}\nonumber\\
    &&+\left[-R''+\beta'R'+\dot{\beta}\dot{R}e^{-2\alpha+2\beta}\right]\left[2 R e^{-2\alpha-2\beta}\left\{-\ddot{R}+\dot{\alpha}\dot{R}+\alpha'R'e^{2\alpha-2\beta}\right\}-(n-5)e^{-2\beta}\sqrt{\frac{F}{R^{n-5}}}\right]\nonumber\\
    &&+(n-5)e^{-2\alpha}\sqrt{\frac{F}{R^{n-5}}}\left[-\ddot{R}+\dot{\alpha}\dot{R}+\alpha'R'e^{2\alpha-2\beta}\right]=\frac{R^{3}}{2(n-3)(n-4)}\left[ p_{t}+\frac{2}{(n-1)}\eta\sigma - \zeta\theta\right]\\
    \Rightarrow \label{R0101}R_{0101}&=& \sqrt{\frac{R^{n-5}}{F}}\frac{R^{2}e^{2\alpha+2\beta}}{2(n-3)(n-4)}\left[p_{t}+\frac{2}{(n-1)}\eta\sigma-\zeta\theta\right]+\frac{(n-5)(n-6)}{4R^{2}}\sqrt{\frac{F}{R^{n-5}}}e^{2\alpha+2\beta}\nonumber\\
    &&-\frac{e^{2\alpha+2\beta}}{R}\sqrt{\frac{R^{n-5}}{F}}\left[2 R e^{-2\alpha-2\beta}\left\{-\ddot{R}+\dot{\alpha}\dot{R}+\alpha'R'e^{2\alpha-2\beta}\right\}-(n-5)e^{-2\beta}\sqrt{\frac{F}{R^{n-5}}}\right]\nonumber\\
    &&.\left[-R''+\beta'R'+\dot{\beta}\dot{R}e^{-2\alpha+2\beta}\right]-\frac{(n-5)}{R}e^{2\beta}\left[-\ddot{R}+\dot{\alpha}\dot{R}+\alpha'R'e^{2\alpha-2\beta}\right]
\end{eqnarray}
Now using $H_{00}$ component,
\begin{eqnarray}
    H_{00}&=&\frac{2e^{-4\beta-2\alpha}(n-2)(n-3)(n-4)}{R^{4}}\bigg[-R{R'}^{3}\beta'e^{4\alpha}-R\dot{R}{R'}^{2}\dot{\beta}e^{2\beta+2\alpha}+R{\dot{R}}^{2}\beta'R'e^{2\beta+2\alpha}+R{\dot{R}}^{3}\dot{\beta}e^{4\beta}\nonumber\\
    &&+\frac{(n-5)}{4}{R'}^{4}e^{4\alpha}+R{R'}^{2}R''e^{4\alpha}-\frac{(n-5)}{2}{\dot{R}}^{2}{R'}^{2}e^{2\beta+2\alpha}-R{\dot{R}}^{2}R''e^{2\beta+2\alpha}+\frac{(n-5)}{4}{\dot{R}}^{4}e^{4\beta}\nonumber\\
    &&+R\beta'R'e^{2\beta+4\alpha}+R\dot{\beta}\dot{R}e^{4\beta+2\alpha}-\frac{(n-5)}{2}{R'}^{2}-RR''e^{2\beta+4\alpha}+\frac{(n-5)}{2}{\dot{R}}^{2}e^{4\beta+2\alpha}\nonumber\\
    &&+\frac{(n-5)}{4}e^{4\alpha+4\beta}\bigg]=\rho e^{2\alpha}
\end{eqnarray}
again using equation (\ref{mass function}),
\begin{equation}\label{R1212}
    R_{1212}=R\beta'R'-RR''+R\dot{\beta}\dot{R}e^{2\beta-2\alpha}=\sqrt{\frac{R^{n-5}}{F}}e^{2\beta}\left[\frac{\rho R^{4}}{2(n-2)(n-3)(n-4)}-\frac{(n-5)}{4}\frac{F}{R^{n-5}}\right]
\end{equation}
Now taking $H_{11}$
\begin{eqnarray}
    H_{11}&=&\frac{2(n-2)(n-3)(n-4)}{R^{4}}\bigg[-R{R'}^{3}\alpha'e^{-2\beta}+R{\dot{R}}^{2}R'\alpha'e^{-2\alpha}+R{\dot{R}}^{3}\dot{\alpha}e^{-4\alpha+2\beta}-R\dot{R}\dot{\alpha}{R'}^{2}e^{-2\alpha}\nonumber\\
    &&-\frac{(n-5)}{4}{R'}^{4}e^{-2\beta}-\frac{(n-5)}{4}{\dot{R}}^{4}e^{-4\alpha+2\beta}-R{\dot{R}}^{2}\ddot{R}e^{-4\alpha+2\beta}+\frac{(n-5)}{4}{\dot{R}}^{2}{R'}^{2}e^{-2\alpha}+R{R'}^{2}\ddot{R}e^{-2\alpha}\nonumber\\
    &&+RR'\alpha'+R\dot{\alpha}\dot{R}e^{2\beta-2\alpha}-\frac{(n-5)}{2}{\dot{R}}^{2}e^{2\beta-2\alpha}-R\ddot{R}e^{2\beta-2\alpha}+\frac{(n-5)}{2}{R'}^{2}+\frac{(n-5)}{4}e^{2\beta}\bigg]\nonumber\\
    &&=e^{2\beta}\left[p_{r}-\frac{2(n-2)}{(n-1)}\eta\sigma-\zeta\theta\right]   
\end{eqnarray}
using equation (\ref{mass function}),
\begin{equation}\label{R0202}
    R_{0202}=R\dot{\alpha}\dot{R}-R\ddot{R}+R\alpha'R'e^{2\alpha-2\beta}=\sqrt{\frac{R^{n-5}}{F}}e^{2\alpha}\left[\frac{R^{4}\left[p_{r}-\frac{2(n-2)}{(n-1)}\eta\sigma-\zeta\theta\right]}{2(n-2)(n-3)(n-4)}+\frac{(n-5)}{4}\frac{F}{R^{n-5}}\right]
\end{equation}
Again using equation (\ref{R1212}) and equation (\ref{R0202}) into equation (\ref{R0101}), we get,
\begin{eqnarray}
    R_{0101}&=&\sqrt{\frac{R^{n-5}}{F}}\frac{R^{2}e^{2\alpha+2\beta}}{2(n-3)(n-4)}\left[p_{t}+\frac{2}{(n-1)}\eta\sigma-\zeta\theta\right]-\frac{(n-5)(n-3)}{8R^{2}}e^{2\alpha+2\beta}\sqrt{\frac{F}{R^{n-5}}}\nonumber\\
    &&-\frac{(n-5)R^{2}e^{2\alpha+2\beta}}{4(n-2)(n-3)(n-4)}\sqrt{\frac{R^{n-5}}{F}}\left[p_{r}-\frac{2(n-2)}{(n-1)}\eta\sigma-\zeta\theta\right]+\frac{(n-5)\rho R^{2}e^{2\alpha+2\beta}}{4(n-2)(n-3)(n-4)}\sqrt{\frac{R^{n-5}}{F}}\nonumber\\
    &&-\frac{\rho R^{6}e^{2\alpha+2\beta}}{2[(n-2)(n-3)(n-4)]^{2}}\left[\frac{R^{n-5}}{F}\right]^{\frac{3}{2}}\left[p_{r}-\frac{2(n-2)}{(n-1)}\eta\sigma-\zeta\theta\right]
\end{eqnarray}
and
\begin{equation}
    R_{2323}=R^{2}\sin^{2}\theta\sqrt{\frac{F}{R^{n-5}}}
\end{equation}
Now Ricci tensors are,
\begin{eqnarray}
    R_{00}&=&g^{11}R_{0101}+(n-2)g^{22}R_{2020}\nonumber\\
    \Rightarrow R_{00}&=& e^{2\alpha}\bigg[\sqrt{\frac{R^{n-5}}{F}}\frac{R^{2}}{2(n-3)(n-4)}\left[p_{t}+\frac{2}{(n-1)}\eta\sigma-\zeta\theta\right]+\frac{(n-5)\rho R^{2}}{4(n-2)(n-3)(n-4)}\sqrt{\frac{R^{n-5}}{F}}\nonumber\\
    &&+\frac{(n+1)R^{2}}{4(n-2)(n-3)(n-4)}\sqrt{\frac{R^{n-5}}{F}}\left[p_{r}-\frac{2(n-2)}{(n-1)}\eta\sigma-\zeta\theta\right]+\frac{(n-1)(n-5)}{8R^{2}}\sqrt{\frac{F}{R^{n-5}}}\nonumber\\
    &&-\frac{\rho R^{6}}{2[(n-2)(n-3)(n-4)]^{2}}\left[\frac{R^{n-5}}{F}\right]^{\frac{3}{2}}\left[p_{r}-\frac{2(n-2)}{(n-1)}\eta\sigma-\zeta\theta\right]\bigg]
\end{eqnarray}
\begin{eqnarray}
    R_{11}&=&g^{00}R_{1010}+(n-2)g^{22}R_{1212}\nonumber\\
    \Rightarrow R_{11}&=& e^{2\beta}\bigg[-\sqrt{\frac{R^{n-5}}{F}}\frac{R^{2}}{2(n-3)(n-4)}\left[p_{t}+\frac{2}{(n-1)}\eta\sigma-\zeta\theta\right]+\frac{(n+1)\rho R^{2}}{4(n-2)(n-3)(n-4)}\sqrt{\frac{R^{n-5}}{F}}\nonumber\\
    &&+\frac{(n-5)R^{2}}{4(n-2)(n-3)(n-4)}\sqrt{\frac{R^{n-5}}{F}}\left[p_{r}-\frac{2(n-2)}{(n-1)}\eta\sigma-\zeta\theta\right]-\frac{(n-1)(n-5)}{8R^{2}}\sqrt{\frac{F}{R^{n-5}}}\nonumber\\
    &&+\frac{\rho R^{6}}{2[(n-2)(n-3)(n-4)]^{2}}\left[\frac{R^{n-5}}{F}\right]^{\frac{3}{2}}\left[p_{r}-\frac{2(n-2)}{(n-1)}\eta\sigma-\zeta\theta\right]\bigg]
\end{eqnarray}
\begin{eqnarray}
    R_{22}&=&g^{00}R_{0202}+g^{11}R_{1212}+(n-3)g^{33}R_{3232}\nonumber\\
    \Rightarrow R_{22}&=& \frac{\rho R^{4}}{2(n-2)(n-3)(n-4)}\sqrt{\frac{R^{n-5}}{F}}-\frac{R^{4}}{2(n-2)(n-3)(n-4)}\sqrt{\frac{R^{n-5}}{F}}\left[p_{r}-\frac{2(n-2)}{(n-1)}\eta\sigma-\zeta\theta\right]\nonumber\\
    &&+\frac{(n-1)}{2}\sqrt{\frac{F}{R^{n-5}}}
\end{eqnarray}
\begin{eqnarray}
    R_{s}&=&g^{00}R_{00}+g^{11}R_{11}+(n-2)g^{22}R_{22}\nonumber\\
    \Rightarrow R_{s}&=& -\sqrt{\frac{R^{n-5}}{F}}\frac{R^{2}}{(n-3)(n-4)}\left[p_{t}+\frac{2}{(n-1)}\eta\sigma-\zeta\theta\right]+\frac{(n+1)\rho R^{2}}{2(n-2)(n-3)(n-4)}\sqrt{\frac{R^{n-5}}{F}}\nonumber\\
    &&-\frac{(n+1)R^{2}}{2(n-2)(n-3)(n-4)}\sqrt{\frac{R^{n-5}}{F}}\left[p_{r}-\frac{2(n-2)}{(n-1)}\eta\sigma-\zeta\theta\right]+\frac{(n-1)(n+1)}{4R^{2}}\sqrt{\frac{F}{R^{n-5}}}\nonumber\\
    &&+\frac{\rho R^{6}}{[(n-2)(n-3)(n-4)]^{2}}\left[\frac{R^{n-5}}{F}\right]^{\frac{3}{2}}\left[p_{r}-\frac{2(n-2)}{(n-1)}\eta\sigma-\zeta\theta\right]
\end{eqnarray}



\begin{thebibliography}{99}

\bibitem{Hawking_Ellis}
S.W. Hawking and  G.F.R. Ellis, The Large Scale Structure of Spacetime,
Cambridge University Press, Cambridge 1975.
\bibitem{Wald}
Robert M. Wald, General Relativity, Univ. of Chicago Press, 1984. 
\bibitem{Penrose:1964wq} 
  R.~Penrose,
  Phys.\ Rev.\ Lett.\  {\bf 14}, 57 (1965).
 \bibitem{Penrose:1969pc} 
  R.~Penrose,
  Riv.\ Nuovo Cim.\  {\bf 1}, 252 (1969)
  [Gen.\ Rel.\ Grav.\  {\bf 34}, 1141 (2002)].
\bibitem{OS}
 J. R. Oppenheimer and H. Snyder, Phys. Rev. {\bf 56}, 455 (1939). 
\bibitem{SD}
 S. Datt, Zs. f. Phys. {\bf 108}, 314 (1938).
\bibitem{Landau_Lifshitz}
L.D. Landau and E.M. Lifshitz, The Classical Theory of Fields, Pergamon Press, 1975. 
 \bibitem{Ori}
 A.~Ori and T.~Piran,
  Phys.\ Rev.\ Lett.\  {\bf 59}, 2137 (1987).
\bibitem{joshi}
P.S. Joshi, Gravitational Collapse and Spacetime Singularities, Cambridge Univ. Press, Cambridge,
 2007.
\bibitem{Joshi:2012mk} 
  P.~S.~Joshi and D.~Malafarina,
  Int.\ J.\ Mod.\ Phys.\ D {\bf 20}, 2641 (2011).
\bibitem{Clarke}  C. J. S. Clarke, The Analysis of Spacetime Singularities,
Cambridge Univ. Press,(1993).
\bibitem{Shapiro:1991zza}
S.~L.~Shapiro and S.~A.~Teukolsky,
Phys. Rev. Lett. \textbf{66}, 994-997 (1991).
\bibitem{Dwivedi:1992fh}
I.~H.~Dwivedi and P.~S.~Joshi,
Class. Quant. Grav. \textbf{9}, L69-L75 (1992).
\bibitem{Ghosh:2001fb}
S.~G.~Ghosh and A.~Beesham,
Phys. Rev. D \textbf{64}, 124005 (2001).
\bibitem{Harada:2001nj}
T.~Harada, H.~Iguchi and K.~i.~Nakao,
Prog. Theor. Phys. \textbf{107}, 449-524 (2002).
\bibitem{Joshi:2004tb}
P.~S.~Joshi, R.~Goswami and N.~Dadhich,
Phys. Rev. D \textbf{70}, 087502 (2004).

\bibitem{Hawking:1972qk}
S.~W.~Hawking,
Commun. Math. Phys. \textbf{25}, 167-171 (1972).
\bibitem{Sotiriou:2011dz}
T.~P.~Sotiriou and V.~Faraoni,
Phys. Rev. Lett. \textbf{108}, 081103 (2012).
\bibitem{Brown:2018hym}
P.~J.~Brown, C.~J.~Fewster and E.~A.~Kontou,
Gen. Rel. Grav. \textbf{50}, no.10, 121 (2018).
\bibitem{Lovelock:1971yv}
D.~Lovelock,
J. Math. Phys. \textbf{12}, 498-501 (1971).
\bibitem{Lanczos:1938sf}
C.~Lanczos,
Annals Math. \textbf{39}, 842-850 (1938).
\bibitem{Dadhich:2012cv}
N.~Dadhich, S.~G.~Ghosh and S.~Jhingan,
Phys. Lett. B \textbf{711}, 196-198 (2012).
\bibitem{Dadhich:2015lra}
N.~Dadhich,
Eur. Phys. J. C \textbf{76}, no.3, 104 (2016).
\bibitem{Camanho:2015hea}
X.~O.~Camanho and N.~Dadhich,
Eur. Phys. J. C \textbf{76}, no.3, 149 (2016).

\bibitem{Kothawala:2009kc}
D.~Kothawala and T.~Padmanabhan,
Phys. Rev. D \textbf{79}, 104020 (2009).
\bibitem{Chakraborty:2014rga}
S.~Chakraborty and T.~Padmanabhan,
Phys. Rev. D \textbf{90}, no.12, 124017 (2014).
\bibitem{Chakraborty:2016qbw}
S.~Chakraborty and N.~Dadhich,
Eur. Phys. J. C \textbf{78}, no.1, 81 (2018).
\bibitem{Gannouji:2019gnb}
R.~Gannouji, Y.~Rodr\'\i{}guez Baez and N.~Dadhich,
Phys. Rev. D \textbf{100}, no.8, 084011 (2019).
\bibitem{Dadhich:2016fku}
N.~Dadhich and S.~Chakraborty,
Phys. Rev. D \textbf{95}, no.6, 064059 (2017).
\bibitem{Dadhich:2016wtb}
N.~Dadhich, S.~Hansraj and B.~Chilambwe,
Int. J. Mod. Phys. D \textbf{26}, no.06, 1750056 (2016).
\bibitem{Shaymatov:2020byu}
S.~Shaymatov and N.~Dadhich,
JCAP \textbf{10}, 060 (2022).
\bibitem{Dadhich:2017zdi}
N.~Dadhich, A.~Molina and J.~M.~Pons,
Phys. Rev. D \textbf{96}, no.8, 084058 (2017).
\bibitem{Molina:2016xeu}
A.~Molina, N.~Dadhich and A.~Khugaev,
Gen. Rel. Grav. \textbf{49}, no.7, 96 (2017).
\bibitem{Singha:2023lum}
C.~Singha and S.~Biswas,
Phys. Rev. D \textbf{109}, no.2, 2 (2024).
\bibitem{Dadhich:2013bya}
N.~Dadhich, S.~G.~Ghosh and S.~Jhingan,
Phys. Rev. D \textbf{88}, 084024 (2013).

 \bibitem{Booth:2005ng} 
  I.~Booth, L.~Brits, J.~A.~Gonzalez and C.~Van Den Broeck,
  Class.\ Quant.\ Grav.\  {\bf 23}, 413 (2006).
\bibitem{Chatterjee:2020khj}
A.~Chatterjee, A.~Ghosh and S.~Jaryal,
Phys. Rev. D \textbf{102}, no.6, 064048 (2020).
\bibitem{Chatterjee:2021zre}
A.~Chatterjee, A.~Ghosh and S.~C.~Jaryal,
Phys. Rev. D \textbf{106}, no.4, 044049 (2022).
\bibitem{Jaryal:2022rzd}
S.~C.~Jaryal and A.~Chatterjee,
Phys. Dark Univ. \textbf{39}, 101171 (2023).
\bibitem{Chatterjee:2024egb}
A.~Chatterjee, S.~C.~Jaryal and A.~Kumar,
Phys. Rev. D \textbf{110}, no.10, 104059 (2024).
\bibitem{Ashtekar:2004cn} 
  A.~Ashtekar and B.~Krishnan,
  Living Rev.\ Rel.\  {\bf 7}, 10 (2004).
\bibitem{Raviteja:2021dgy}
K.~Raviteja, A.~Haque and S.~Gutti,
Phys. Rev. D \textbf{103}, no.12, 124005 (2021).
 \bibitem{gutti2}
  A. Pathak, K. Raviteja, S. Bhattacharya, S. Gutti, Phys.\ Rev.\ D {\bf 109}, 084062 (2024).
\bibitem{Maeda:2005ci}
H.~Maeda,
Class. Quant. Grav. \textbf{23}, 2155 (2006).
\bibitem{Cai:2006pq}
R.~G.~Cai and N.~Ohta,
Phys. Rev. D \textbf{74}, 064001 (2006).
\bibitem{Garraffo:2008hu}
C.~Garraffo and G.~Giribet,
Mod. Phys. Lett. A \textbf{23}, 1801-1818 (2008).
\bibitem{Mukherjee:2020lld}
S.~Mukherjee and N.~Dadhich,
Eur. Phys. J. C \textbf{81}, no.5, 458 (2021).
\bibitem{Mukherjee:2021erg}
S.~Mukherjee and N.~Dadhich,
Eur. Phys. J. C \textbf{82}, no.4, 302 (2022).
\bibitem{Dialektopoulos:2023qda}
K.~F.~Dialektopoulos, D.~Malafarina and N.~Dadhich,
Phys. Rev. D \textbf{108}, no.4, 044080 (2023).
 
 
 
          
          
\end{thebibliography}
\end{document}